\documentclass[a4paper,11pt]{article}
\usepackage[a4paper, margin=1in]{geometry}  
\usepackage{setspace}  
\usepackage{graphicx}  
\usepackage{amsmath, amssymb}  
\usepackage{booktabs}  
\usepackage{array}  
\usepackage{times}  
\usepackage[T1]{fontenc}  
\usepackage[utf8]{inputenc}  
\usepackage{lmodern}  
\usepackage{cancel}  
\usepackage{slashed}
\usepackage{hyperref}  
\usepackage{listings}  
\usepackage{color}     
\usepackage{cite}      
\usepackage{float}
\usepackage{multirow}
\usepackage[xcdraw]{xcolor}
\usepackage[utf8]{inputenc}  
\usepackage{graphicx}  
\usepackage{hyperref}  
\hypersetup{unicode=true}  
\usepackage{pdfpages}  
\usepackage{float}  

\providecommand{\openone}{\leavevmode\hbox{\small1\kern-3.8pt\normalsize1}}

\newcommand{\gm}{\gamma^\mu}
\newcommand{\Wm}{W_{\mu}}
\newcommand{\Zm}{Z_{\mu}}
\newcommand{\ts}{$(T)$}
\newcommand{\bs}{$(B)$}
\newcommand{\xt}{$(X\,T)$}
\newcommand{\tb}{$(T\,B)$}
\newcommand{\by}{$(B\,Y)$}
\newcommand{\xtb}{$(X\,T\,B)$}
\newcommand{\tby}{$(T\,B\,Y)$}

\newcommand{\slx}{s_L}
\newcommand{\slu}{s_L^u}
\newcommand{\sld}{s_L^d}
\newcommand{\clx}{c_L}
\newcommand{\clu}{c_L^u}
\newcommand{\cld}{c_L^d}
\newcommand{\srx}{s_R}
\newcommand{\sru}{s_R^u}
\newcommand{\srd}{s_R^d}
\newcommand{\crx}{c_R}
\newcommand{\cru}{c_R^u}
\newcommand{\crd}{c_R^d}

\newcommand{\sqt}{\sqrt{2}}

\newcommand{\RE}{\text{Re}}

\usepackage{courier}  
\usepackage{adjustbox}

\begin{document}
\thispagestyle{empty}

\def\thefootnote{\fnsymbol{footnote}}

\begin{center}
    \Large\boldmath\textbf{
        Vector-Like Quarks at the LHC: A Unified Perspective from ATLAS and CMS Exclusion Limits}
    \unboldmath
\end{center}
\vspace{-0.5cm}
\begin{center}
    R. Benbrik$^a$  \footnote{
    	\href{mailto:r.benbrik@uca.ac.ma}{r.benbrik@uca.ac.ma}
    }, M. Boukidi$^a$ \footnote{
    \href{mailto:mohammed.boukidi@ced.uca.ma}{mohammed.boukidi@ced.uca.ma}
}, M. Ech-chaouy$^a$  \footnote{
\href{mailto:m.echchaouy.ced@uca.ac.ma}{m.echchaouy.ced@uca.ac.ma}}
, S. Moretti$^{b,c}$  \footnote{
	\href{mailto:stefano.moretti@physics.uu.se}{stefano.moretti@physics.uu.se},	\href{mailto:s.moretti@soton.ac.uk}{s.moretti@soton.ac.uk}}
, K. Salime$^a$  \footnote{
\href{mailto:k.salime.ced@uca.ac.ma}{k.salime.ced@uca.ac.ma}}
,  Q.S. Yan$^{d,e}$  \footnote{\href{mailto:yanqishu@ucas.ac.cn}{yanqishu@ucas.ac.cn}}
\vspace{0.5cm}

    \textsl{\small
        $^a$Polydisciplinary Faculty, Laboratory of Physics, Energy, Environment, and Applications, Cadi Ayyad University, Sidi Bouzid, B.P. 4162, Safi, Morocco.\\
        $^b$Department of Physics \& Astronomy, Uppsala University, Box 516, SE-751 20 Uppsala, Sweden.\\
        $^c$School of Physics \& Astronomy, University of Southampton, Southampton, SO17 1BJ, United Kingdom.\\
        $^d$Center for Future High Energy Physics, Chinese Academy of Sciences, Beijing 100049, P.R. China.\\   
        $^e$School of Physics Sciences, University of Chinese Academy of Sciences, Beijing 100039, P.R. China.
    }

\end{center}
\vspace{0.2cm}

\renewcommand{\thefootnote}{\arabic{footnote}}
\setcounter{footnote}{0}

\begin{abstract}
In this work, we present a comprehensive review of the most up-to-date exclusion limits on Vector-Like Quarks (VLQs) derived from ATLAS and CMS data at the Large Hadron Collider (LHC). Our analysis encompasses both pair and single production modes, systematically comparing results from the two collaborations to identify and employ the most stringent bounds at each mass point. We evaluate the excluded parameter space for VLQs under singlet, doublet, and triplet representations. For top-like VLQs ($T$), the exclusion limits rule out masses up to 1.49 TeV in singlet scenarios, while single production constrains the mixing parameter $\kappa$ to values below 0.26 at $m_T \sim 1.5$ TeV and up to 0.42 for $m_T \sim 2$ TeV. For bottom-like VLQs ($B$), the strongest exclusion limits from pair production exclude masses up to 1.52 TeV in doublet configurations, with single production constraining $\kappa$ values between 0.2 and 0.7 depending on the mass. For exotic VLQs, such as $X$ and $Y$, pair production excludes masses up to 1.46 TeV and 1.7 TeV, respectively. The constraints on $\kappa$ from these analyses become increasingly restrictive at higher masses, reflecting the enhanced sensitivity of single production channels in this regime. For $X$, $\kappa$ is constrained below 0.16 for masses between 0.8 and 1.6 TeV and further tightens to $\kappa < 0.2$ as the mass approaches 1.8 TeV. Similarly, for $Y$, $\kappa$ values are constrained below 0.26 around $m_Y \sim 1.7$ TeV, with exclusions gradually relaxing at higher masses. These exclusion regions, derived from the most stringent LHC search results, offer a unified and up-to-date perspective on VLQ phenomenology. The results were computed using \texttt{VLQBounds}, a new Python-based tool specifically developed for this purpose.
\end{abstract}

\clearpage

\tableofcontents

\clearpage

\section{Introduction}          
The SM has been remarkably successful in explaining a wide range of particle physics phenomena, as evidenced by its predictions and experimental validation through key discoveries such as the Higgs boson~\cite{ATLAS:2012yve, CMS:2012qbp}. However, it is widely considered a low-energy approximation of a more fundamental theory. As the search for new physics intensifies, experimental data continues to impose tighter constraints on potential deviations from SM predictions, which are critical in guiding us toward new discoveries.

One of the most natural and intriguing extensions of the SM involves the introduction of VLQs \cite{Aguilar-Saavedra:2013qpa,Buchkremer:2013bha,Fuks:2016ftf,Alves:2023ufm}. These exotic fermions differ from SM quarks in that they possess both left- and right-handed components that transform identically under the SM gauge group $SU(3)_c \times SU(2)_L \times U(1)_Y$. The addition of VLQs enriches the theoretical framework, leading to novel phenomena that can be explored at the LHC and future colliders. VLQs can exist as singlets, doublets or triplets under $SU(2)_L$ and interact with SM quarks through Yukawa couplings a Higgs fields~\cite{delAguila:2000aa}. Their theoretical appeal is further enhanced by their appearance in several beyond-the-Standard-Model (BSM) scenarios. In extra-dimensional models, such as the Randall-Sundrum warped geometry~\cite{Randall:1999ee}, VLQs emerge as Kaluza-Klein excitations of quarks when the extra dimensions are compactified~\cite{Carena:2007tn, Gopalakrishna:2011ef}. VLQs also appear naturally in Grand Unified Theories (GUTs), such as those based on an $E_6$ symmetry~\cite{Hewett:1988xc}, and in other models like Little Higgs~\cite{Arkani-Hamed:2002ikv, Schmaltz:2005ky}  as well as composite Higgs frameworks\cite{Dobrescu:1997nm, Chivukula:1998wd, He:2001fz, Hill:2002ap, Agashe:2004rs, Contino:2006qr, Barbieri:2007bh, Anastasiou:2009rv}.  In addition to these complete SM extensions, the phenomenology of VLQs has also been studied extensively within the 2-Higgs-Doublet Model (2HDM) \cite{Arhrib:2024dou, Arhrib:2024tzm, Arhrib:2024mbq, Benbrik:2022kpo,Abouabid:2023mbu,Benbrik:2024hsf,Arhrib:2024nbj,Benbrik:2023xlo,Benbrik:2024bxt, Dermisek:2019vkc,Cingiloglu:2023ylm}.

VLQs possess gauge-invariant mass terms, allowing their masses to be significantly larger than the Electro-Weak (EW) scale, distinguishing these from the now-excluded fourth generation of chiral quarks. Their mixing with SM quarks introduces small deviations from the unitarity of the Cabibbo-Kobayashi-Maskawa (CKM) matrix, the latter being proportional to the ratio $m/M$, where $m$ is the EW scale and $M$ is the VLQ mass. This mixing also generates Flavor-Changing Neutral Currents (FCNCs) mediated by $Z$ and  Higgs boson currents, which are however suppressed by the same $m/M$ ratio. The presence of VLQs also influences flavor physics, weak processes and meson mixing. These distinctive features make VLQs promising candidates for addressing unresolved issues of the SM, such as the CKM unitarity problem~\cite{Seng:2018yzq}, and they may also provide explanations for recent anomalies like the $W$ boson mass measurement by CDF~\cite{CDF:2022hxs}.

At the LHC, searches for VLQs are primarily conducted in two production modes: pair production and single production. Pair production, driven by QCD interactions, is largely model-independent, as its cross-section  simply depend upon the VLQ mass. Single production, in contrast, is mediated by EW interactions, making it more sensitive to the specific couplings between VLQs and SM quarks.

To date, the ATLAS and CMS collaborations have conducted a total of 61 studies~\cite{ATLAS:2012qe, ATLAS:2012tkh, ATLAS:2012wxk, ATLAS:2014vpn, ATLAS:2015ktd, ATLAS:2015uaw, ATLAS:2015vzd, ATLAS:2016btu, ATLAS:2016ovj, ATLAS:2016scx, ATLAS:2016seq, ATLAS:2016sno, ATLAS:2017nap, ATLAS:2017vdo, ATLAS:2018alq, ATLAS:2018cjd, ATLAS:2018cye, ATLAS:2018dyh, ATLAS:2018mpo, ATLAS:2018tnt, ATLAS:2018uky, ATLAS:2018ziw, ATLAS:2022hnn, ATLAS:2022ozf, ATLAS:2022tla, ATLAS:2023bfh, ATLAS:2023ixh, ATLAS:2023pja, ATLAS:2024gyc, ATLAS:2024kgp, ATLAS:2024xdc, ATLAS:2024xne, CMS:2012ab, CMS:2012mir, CMS:2013hwy, CMS:2015hyy, CMS:2015jwh, CMS:2015lzl, CMS:2016edj, CMS:2016ete, CMS:2016jce, CMS:2017fpk, CMS:2017gsh, CMS:2017ked, CMS:2017voh, CMS:2017ynm, CMS:2018dcw, CMS:2018kcw, CMS:2018ubm, CMS:2018wpl, CMS:2018zkf, CMS:2019afi, CMS:2019eqb, CMS:2020ttz, CMS:2021mku, CMS:2022fck, CMS:2022yxp, CMS:2023agg, CMS:2024bni, CMS:2024qdd, CMS:2024xbc} investigating VLQ production at the LHC. Among these, 42.4\% focus on pair production in ATLAS and 22.2\% in CMS, while 18.2\% and 17.2\% address single production in ATLAS and CMS, respectively. The strong emphasis on pair production arises from its robustness and model independence, although it is limited by significant phase space suppression for VLQ masses in the TeV range. Conversely, single production provides complementary insights by probing the electroweak couplings of VLQs, which are essential for understanding their multiplet representations. Additionally, single production becomes increasingly relevant for exploring heavy VLQs, as its cross-section decreases more slowly with increasing mass compared to pair production.

In the, for short, SM+VLQ framework, VLQ decays are dominated by channels involving heavy quarks and EW  bosons, yielding distinct final states that facilitate identification in collider experiments. Top-like VLQs (VLTs) $T$ predominantly decay via $T \rightarrow Wb$, $T \rightarrow Zt$ and $T \rightarrow ht$, while bottom-like VLQs (VLBs) $B$ decay through analogous channels such as $B \rightarrow Wt$, $B \rightarrow Zb$ and $B \rightarrow hb$. By contrast, the exotic states $X$ and $Y$, due to their unique electric charges, exhibit single decay modes within the SM+VLQ framework: $X \rightarrow W^+ t$ and $Y \rightarrow W^- b$, respectively.  These decay modes lead to final states rich in vector bosons and third-generation quarks, providing distinct signatures for experimental searches and enabling a comprehensive study of VLQ multiplets.

\begin{figure}[H]
	\centering
	\includegraphics[scale=0.65]{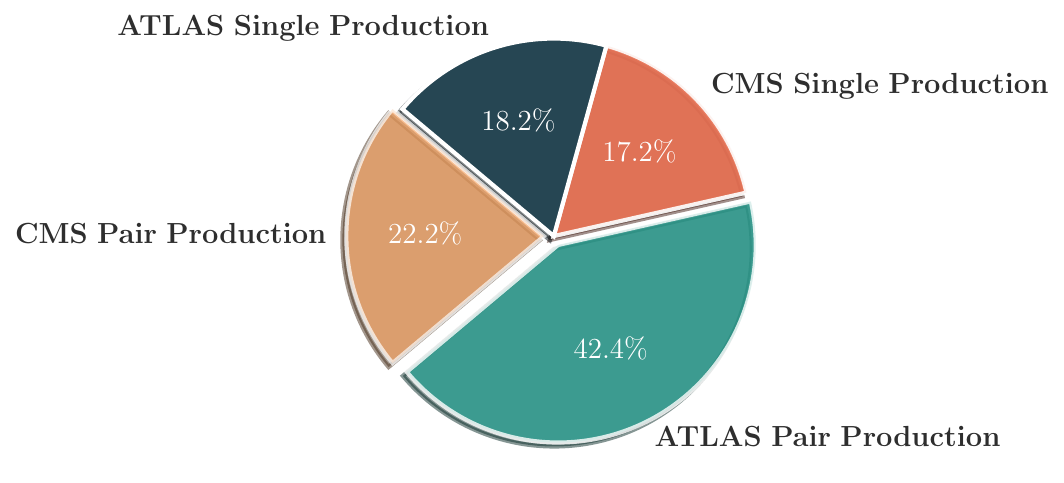}
	\caption{Distribution of VLQ production studies by ATLAS and CMS at the LHC, covering pair and single production analyses.}
	\label{fig:1}
\end{figure}

Fig.~\ref{fig:1} illustrates the distribution of VLQ production studies by ATLAS and CMS, highlighting the focus of each collaboration on pair and single production analyses.

A recent study~\cite{Banerjee:2024zvg} provides an extensive analysis of VLQ searches at the LHC, concentrating on exclusion limits from both ATLAS and CMS and further extending its scope to include recast results for VLQ decays into BSM particles, such as new scalar, pseudoscalar and charged Higgs bosons. In contrast, our work provides a comprehensive examination of direct VLQ searches conducted by these collaborations, emphasizing the most recent (and typically more constraining)  exclusion limits obtained from both single and pair production modes. Additionally, we present results computed with \texttt{VLQBounds}, a Python-based tool newly developed to systematically evaluate theoretical predictions against these experimental limits, which will be publicly released in the near future.

Our paper is organized as follows. Sect. \ref{sec:framework} provides an overview of the theoretical framework, including a detailed discussion of VLQ production mechanisms at the LHC, focusing on both pair and single production modes. Sect. \ref{sec:result} presents the results of ATLAS and CMS searches for each VLQ type ($T$, $B$, $X$ and $Y$), emphasizing the distinct exclusion limits derived from these studies. Sect. \ref{sec:Global} integrates the strongest $95\%$ Confidence Level (CL) limits from ATLAS and CMS into a global exclusion analysis. Sect. \ref{sec:conclusion} concludes with a summary of our main findings. Finally, the Appendix includes analytical expressions for the couplings of both VLQs and SM quarks to the gauge and Higgs boson, along with a guide to the coupling notation conventions used throughout this paper.

\section{Model Framework and Notation}
\label{sec:framework}

In a framework where the scalar sector is limited to ${SU}(2)_L$ doublets, as in the SM, new VLQs that couple to SM quarks via renormalizable interactions can only exist within seven distinct gauge-covariant multiplets, each characterized by specific ${SU}(3)_C \times {SU}(2)_L \times {U}(1)_Y$ quantum numbers~\cite{delAguila:2000aa}: 
\begin{align}
& T_{L,R}^0 \,, \quad B_{L,R}^0 && \text{(singlets)} \,, \notag \\
& (X\,T^0)_{L,R} \,, \quad (T^0\,B^0)_{L,R} \,, \quad (B^0\,Y)_{L,R} && \text{(doublets)} \,, \notag \\
& (X\,T^0\,B^0)_{L,R} \,, \quad (T^0\,B^0\,Y)_{L,R}  && \text{(triplets)} \,.
\end{align}
In this notation, a superscript zero is used to denote weak eigenstates to differentiate them from mass eigenstates (this distinction will be omitted where context makes it clear). The fields $T^0$ and $B^0$ represent quarks with electric charges of $2/3$ and $-1/3$, respectively. Some multiplets also contain VLQs with exotic charges, $X$ ($5/3$) and $Y$ ($-4/3$), and in cases where only one such multiplet is present, the weak and mass eigenstates coincide. For simplicity, we will limit our discussion to extensions of the SM involving only one additional multiplet.

When the $T_{L,R}^0$ quarks of charge $2/3$ are introduced with non-standard isospin assignments, the physical up-type quark mass eigenstates, $u, c, t, T$, may contain contributions from $T_{L,R}^0$, resulting in deviations in their couplings to the $Z$ boson. 
In this scenario, the relationship between the charge $2/3$ weak and mass eigenstates can be expressed via two $2 \times 2$ unitary matrices, $U_{L,R}^u$, parameterized as follows: 
\begin{equation}
\left(\! \begin{array}{c} t_{L,R} \\ T_{L,R} \end{array} \!\right) =
U_{L,R}^u \left(\! \begin{array}{c} t^0_{L,R} \\ T^0_{L,R} \end{array} \!\right)
= \left(\! \begin{array}{cc} \cos \theta_{L,R}^u & -\sin \theta_{L,R}^u e^{i \phi_u} \\ \sin \theta_{L,R}^u e^{-i \phi_u} & \cos \theta_{L,R}^u \end{array}
\!\right)
\left(\! \begin{array}{c} t^0_{L,R} \\ T^0_{L,R} \end{array} \!\right) \,.
\label{ec:mixu}
\end{equation}
Similarly, adding $B_{L,R}^0$ fields with charge $-1/3$ results in four mass eigenstates, $d, s, b, B$. The mixing in the down sector is also described using two $2 \times 2$ unitary matrices, $U_{L,R}^d$: 
\begin{equation}
\left(\! \begin{array}{c} b_{L,R} \\ B_{L,R} \end{array} \!\right)
= U_{L,R}^d \left(\! \begin{array}{c} b^0_{L,R} \\ B^0_{L,R} \end{array} \!\right)
= \left(\! \begin{array}{cc} \cos \theta_{L,R}^d & -\sin \theta_{L,R}^d e^{i \phi_d} \\ \sin \theta_{L,R}^d e^{-i \phi_d} & \cos \theta_{L,R}^d \end{array}
\!\right)
\left(\! \begin{array}{c} b^0_{L,R} \\ B^0_{L,R} \end{array} \!\right) \,.
\label{ec:mixd}
\end{equation}
The complete Lagrangian for the third-generation (chiral) quarks and VLQs in the mass eigenstate basis is provided in  Appendix~\ref{sec:a}. To streamline the notation, we omit the superscripts $u$ and $d$ in the mixing angles $\theta^u_{L,R}$ and $\theta^d_{L,R}$ where only one type of mixing occurs. Additionally, shorthand notations are used: $s_{L,R}^{u,d} \equiv \sin \theta_{L,R}^{u,d}$, $c_{L,R}^{u,d} \equiv \cos \theta_{L,R}^{u,d}$, etc.
This Lagrangian incorporates all the essential elements for phenomenological studies, as follows. \begin{itemize} \item[(i)] Modifications of SM couplings, revealing indirect effects VLQs, can be observed in terms without BSM fields. \item[(ii)] Terms involving one SM quark and one VLQ are crucial for analyzing VLQ production and decay at the LHC. \item[(iii)] Contributions from two VLQs are relevant for oblique corrections and EW Precision Observables (EWPOs). \end{itemize}
The unitary matrices $U_{L,R}^u$ in Eq.(\ref{ec:mixu}) and $U_{L,R}^d$ in Eq.(\ref{ec:mixd}) are defined to ensure that the mass matrices are diagonal in the mass eigenstate basis. In the weak eigenstate basis, the mass terms for the third-generation and heavy quarks are given by:
\begin{eqnarray}
\mathcal{L}_\text{mass} & = & - \left(\! \begin{array}{cc} \bar t_L^0 & \bar T_L^0 \end{array} \!\right)
\left(\! \begin{array}{cc} y_{33}^u \frac{v}{\sqrt 2} & y_{34}^u \frac{v}{\sqrt 2} \\ y_{43}^u \frac{v}{\sqrt 2} & M^0 \end{array} \!\right)
\left(\! \begin{array}{c} t^0_R \\ T^0_R \end{array}
\!\right) \notag \\
& & - \left(\! \begin{array}{cc} \bar b_L^0 & \bar B_L^0 \end{array} \!\right)
\left(\! \begin{array}{cc} y_{33}^d \frac{v}{\sqrt 2} & y_{34}^d \frac{v}{\sqrt 2} \\ y_{43}^d \frac{v}{\sqrt 2} & M^0 \end{array} \!\right)
\left(\! \begin{array}{c} b^0_R \\ B^0_R \end{array}
\!\right) +\text{H.c.} \,,
\label{ec:Lmass}
\end{eqnarray}
Here, $y_{ij}^q$ ($q = u, d$) denote the Yukawa couplings, $v = 246$ GeV represents the Higgs Vacuum Expectation Value (VEV), and $M^0$ is a bare mass term\footnote{It is important to note that this term is not connected to the Higgs mechanism, however, it is gauge-invariant and may appear as a bare mass term in the Lagrangian or arise from a Yukawa coupling to a singlet field that acquires a VEV $v' \gg v$.}. Consequently, the mixing matrices are determined by:
\begin{equation}
U_L^q \, \mathcal{M}^q \, (U_R^q)^\dagger = \mathcal{M}^q_\text{diag} \,, 
\label{ec:diag}
\end{equation}
With $\mathcal{M}^q$ denoting the mass matrices in Eq.~(\ref{ec:Lmass}) and $\mathcal{M}^q_\text{diag}$ their diagonalized forms, these general expressions simplify in specific scenarios. In multiplets where either $T$ or $B$ quarks are absent, the respective $2 \times 2$ mass matrix reduces to the SM quark mass term. Additionally, in multiplets containing both $T$ and $B$ quarks, the bare mass term is identical for the up- and down-type quark sectors. For singlets and triplets one has $y_{43}^q = 0$, whereas for doublets $y_{34}^q=0$. Moreover, for the \xtb\ triplet $y_{34}^d = \sqrt 2 y_{34}^u$ and for the \tby\ triplet, $y_{34}^u = \sqrt 2 y_{34}^d$\footnote{The triplets are expressed in the spherical basis, where the $\sqrt{2}$ factors arise from the correspondence between the Cartesian and spherical components of a rank-1 irreducible tensor operator (vector).}..

The mixing angles in the left- and right-handed sectors are not independent. From the bi-unitary diagonalization of the mass matrix, as shown in Eq.~(\ref{ec:diag}), the following relations are obtained:
\begin{eqnarray}
\tan 2 \theta_L^q & = & \frac{\sqrt{2} |y_{34}^q| v M^0}{(M^0)^2-|y_{33}^q|^2 v^2/2 - |y_{34}^q|^2 v^2/2} \quad \text{(singlets, triplets)} \,, \notag \\
\tan 2 \theta_R^q & = & \ \frac{\sqrt{2}  |y_{43}^q| v M^0}{(M^0)^2-|y_{33}^q|^2 v^2/2 - |y_{43}^q|^2 v^2/2} \quad \text{(doublets)} \,, 
\label{ec:angle1}
\end{eqnarray}
with the relations:
\begin{eqnarray}
\tan \theta_R^q & = & \frac{m_q}{m_Q} \tan \theta_L^q \quad \text{(singlets, triplets)} \,, \notag \\
\tan \theta_L^q & = & \frac{m_q}{m_Q} \tan \theta_R^q \quad \text{(doublets)} \,, 
\label{ec:rel-angle1}
\end{eqnarray}
with $(q,m_q,m_Q) = (u,m_t,m_T)$ and $(d,m_b,m_B)$, one of the mixing angles is typically dominant, particularly in the down sector. Moreover, for triplet representations, the correlations among the off-diagonal Yukawa couplings give rise to specific relations between the mixing angles in the up and down sectors,
\begin{eqnarray}
\sin 2\theta_L^d & = & \sqrt{2} \, \frac{m_T^2-m_t^2}{m_B^2-m_b^2} \sin 2 \theta_L^u \quad \quad (X\,T\,B) \,, \notag \\
\sin 2\theta_L^d & = & \frac{1}{\sqrt{2}} \frac{m_T^2-m_t^2}{m_B^2-m_b^2} \sin 2 \theta_L^u \quad \quad (T\,B\,Y) \,.
\end{eqnarray}
This analysis shows that, for most VLQ multiplets, a single independent mixing angle characterizes the model, except for the \tb\ doublet, which requires two mixing angles. The masses of the heavy quarks deviate from the bare mass $M^0$ due to their mixing with SM quarks. For doublets and triplets, the masses of the different components are correlated, as described by the following relations: \begin{eqnarray} m_X^2 &=& m_T^2 \cos^2 \theta_X + m_t^2 \sin^2 \theta_X \quad \quad \text{for \xt}, \notag\\\ m_T^2 \cos^2 \theta_u + m_t^2 \sin^2 \theta_u &=& m_B^2 \cos^2 \theta_d + m_b^2 \sin^2 \theta_d \quad \quad ~\text{for \tb}, \notag\\ m_Y^2 &=& m_B^2 \cos^2 \theta_Y + m_b^2 \sin^2 \theta_Y \quad \quad \text{for \by}. \end{eqnarray}

These relations demonstrate that the masses of the multiplet components are interdependent, with the heavy quark $T$ typically being heavier than the $X$ quark and the $B$ one being heavier than $Y$. Additionally, $T$ and $B$ may have comparable masses,  depending on the mixing angles. Altogether, most VLQ multiplets can be characterized by a single mixing angle, a heavy quark mass  and, in some cases, a CP-violating phase, which has a minimal effect on the observables relevant to this work. However, for the \tb\ doublet, two independent mixing angles and CP phases are required for a complete parameterization of the model.

\subsection{Pair Production}

Pair production of VLQs is predominantly governed by QCD interactions, rendering the process largely model-independent. The primary production mechanisms at the LHC involve gluon-gluon fusion and quark-antiquark annihilation, leading to the creation of VLQ pairs. The production cross-section depends primarily on the VLQ mass, with minimal dependence on their couplings or decay modes, allowing precise theoretical predictions.

For $T$ and $B$ VLQs, pair production searches focus on final states arising from these decays:
\begin{itemize}
	\item $T \rightarrow Wb$, $T \rightarrow Zt$, $T \rightarrow ht$,
	\item $B \rightarrow Wt$, $B \rightarrow Zb$, $B \rightarrow hb$.
\end{itemize}

Similarly, $X$ and $Y$ VLQs, which have exotic electric charges of ${5}/{3}$ and $-{4}/{3}$ respectively, offer unique signatures due to their distinctive decay modes:
\begin{itemize}
	\item $X \rightarrow W^+ t$,
	\item $Y \rightarrow W^- b$.
\end{itemize}
In essence, 
the $X$ quark decays exclusively into $W^+ t$, leading to final states with Same-Sign (SS)\footnote{In contrast to Opposite-Signed (OS).} $2\ell$ when considering pair production, a signature with relatively low background in the SM. The $Y$ quark decays into $W^- b$, resulting in similar but charge-conjugate final states.

Generally speaking, experimental signatures of VLQ pair production are characterized  by relatively large cross sections due to its QCD nature when compared with their, e.g.,  Vector-Like Leptons (VLLs) with the same mass. Furthermore, according to the available decay channels of the VLQs, multi-weak vector bosons and $b$-jets can be produced, which can  then lead to  final states with multiple leptons, light jets (in addition to the $b$-ones)  and Missing Transverse Energy (MET or $\cancel{E}_T$). 

\begin{figure}[h!]
	\centering
 \centering \includegraphics[height=4cm,width=5cm]{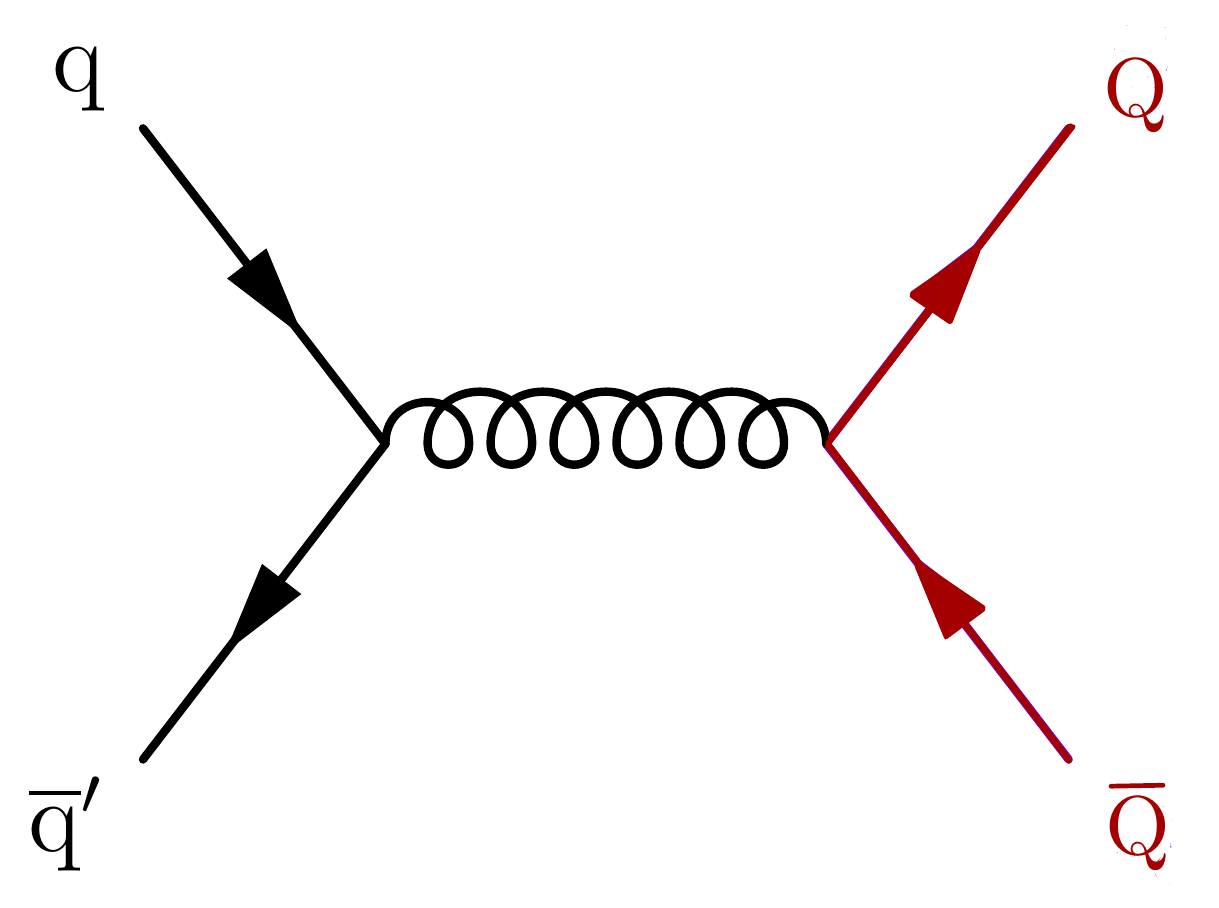}\hspace{1cm} \includegraphics[height=4cm,width=5cm]{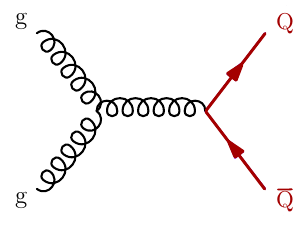}   
	\caption{Feynman diagram for pair production of VLQs via quark-antiquark fusion (left) and gluon-gluon fusion (right) at the LHC.}
	\label{fig:pair_production}
\end{figure}

Representative Feynman diagrams for VLQ pair production via quark-antiquark annihilation and gluon-gluon fusion are shown in Fig.~\ref{fig:pair_production}.

\subsection{Single Production}

Single production of VLQs occurs via EW interactions and is highly sensitive to the specific couplings between VLQs and SM quarks. These couplings arise from mixing between VLQs and third-generation quarks, leading to FCNCs alongside charged currents.

For $T$, $B$, $X$ and $Y$ quarks, single production channels include processes like $pp \rightarrow Qq$, where $Q$ is a VLQ and $q$ is a light SM quark. These processes proceed via the exchange of a $W$, $Z$ or Higgs boson and the VLQ subsequently decays into SM particles according to its decay modes.
{At higher VLQ masses, single production becomes increasingly relevant due to its slower decrease in cross-section compared to pair production \cite{CMS:2024bni}. This distinction arises from reduced phase-space suppression in single production (with respect to pair production) and the growing dominance of quark Parton Distribution Functions (PDFs) (over gluon ones) at higher energy scales. As mentioned, while pair production is driven by QCD interactions and remains independent of VLQ couplings to SM bosons, single production is mediated by EW processes and is strongly influenced by these couplings. If the latter are sufficiently large, single production\footnote{For VLQ masses beyond 1.5 TeV and mixing angles around 10\%, the cross-section for single production can surpass that of pair production. This behavior reflects the dynamics of the processes: while pair production suffers from significant phase-space suppression at high masses, single production benefits from quark-initiated processes that dominate gluon-initiated ones in the PDFs. Additionally, the suppression of $s$-channel contributions at high masses further enhances single production.} can provide significant contributions, not only  at high masses, potentially overtaking pair production as the dominant production mechanism within the energy reach of the LHC. However, if the couplings are too small, single production may be strongly suppressed, even at large VLQ masses, limiting its observability despite the phase space enhancement. The dependence on such couplings allows these processes to be sensitive to scenarios with significant mixing with SM quarks, providing complementary information to pair production analyses. A representative Feynman diagrams for VLQ single production is shown in Fig.~\ref{fig:single_production}.

\begin{figure}[h!]
	\centering
 \centering \includegraphics[height=4.25cm,width=7.cm]{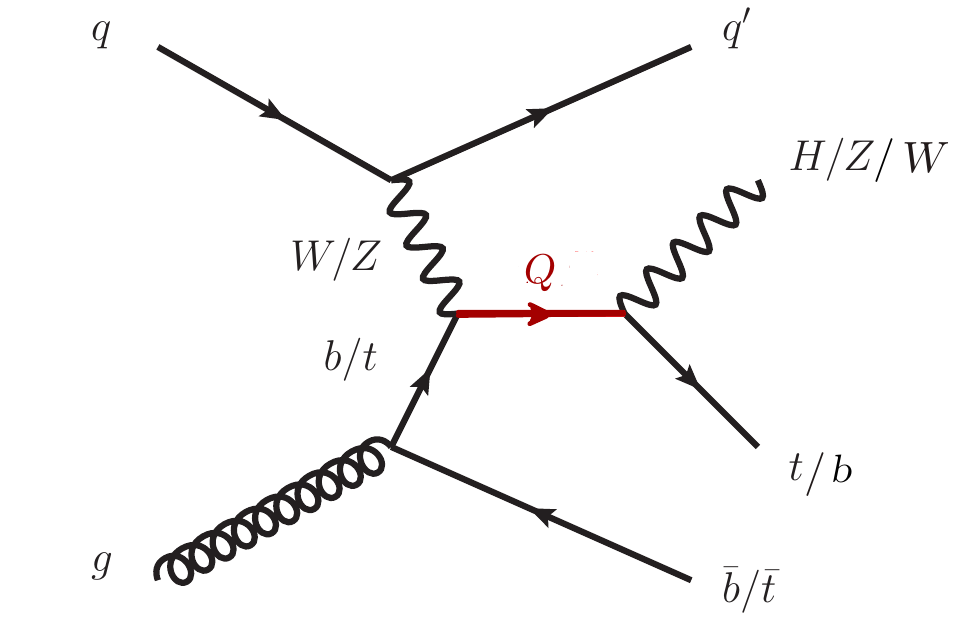}\hspace{1cm} \includegraphics[height=4.25cm,width=5.5cm]{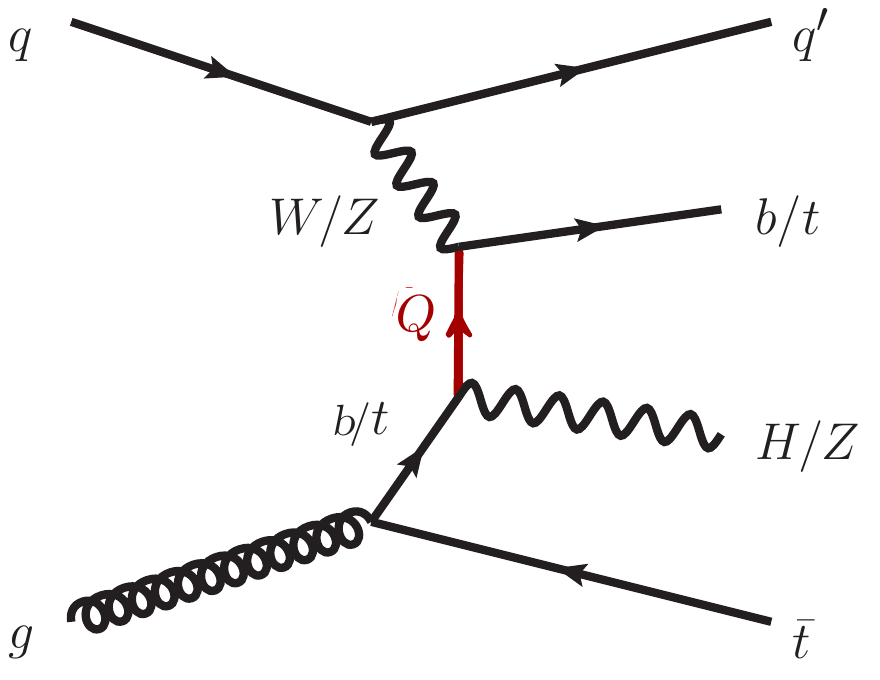} 
	\caption{Feynman diagrams for single production of a VLQ  via EW interactions at the LHC.}
	\label{fig:single_production}
\end{figure}

\section{ATLAS and CMS Search Strategies}
\label{sec:result}

ATLAS and CMS employ diverse strategies at the LHC to probe signals from VLQs. Tables~\ref{tab1} to \ref{tab6} summarize the analyses conducted during Run 1 and Run 2, covering both single and pair production modes. Final states are classified by the presence of $b$-jets, as well as jet and lepton multiplicities, to capture the wide range of expected VLQ signatures.

\noindent Exclusion limits are presented in terms of the coupling parameter $\kappa$, which represents the mixing angles between VLQs ($T$, $B$, $X$, $Y$) and third-generation SM quarks, following the framework in Ref.\cite{Buchkremer:2013bha}. Results from Ref.\cite{CMS:2024bni}, which adopt an alternative parameterization discussed in Ref.\cite{Fuks:2016ftf}, are converted to the $\kappa$ notation of Ref.\cite{Buchkremer:2013bha} to maintain consistency with standard conventions. Appendix~\ref{apB} provides details on the conversion between the theoretical notation used in this study (based on Ref.~\cite{Aguilar-Saavedra:2013qpa}) and the experimental $\kappa$ parameter.

\noindent The exclusion bounds are determined using the standard approach~\cite{Bechtle:2008jh}: 
\begin{eqnarray}
\mathrm{ratio} = \frac{\sigma_{\mathrm{theo}}}{\sigma_{\mathrm{obs}}},
\end{eqnarray}

where a point is excluded if $\mathrm{ratio} \geq 1$. This methodology\footnote{The theoretical cross-sections used in this study are derived from experimental references. Pair production cross-sections are computed at NNLO in QCD using \texttt{top++}~\cite{Czakon:2013goa,Matse:2014,Czakon:2011xx}, while single production cross-sections are calculated at LO in EW interactions under the NWA, following Refs.~\cite{Campbell:2004,Matse:2014,Fuks:2016ftf,Carvalho:2018jkq}.} directly compares theoretical predictions with observed cross-section limits and has been consistently employed in prior VLQ analyses to establish exclusion bounds.
\subsection{VLT Searches}

This section presents the exclusion limits for top-like VLQs (VLTs), derived from both ATLAS and CMS data, and examines these limits for singlet and doublet configurations.

\begin{figure}[H]
	\centering
	\includegraphics[height=7cm,width=16cm]{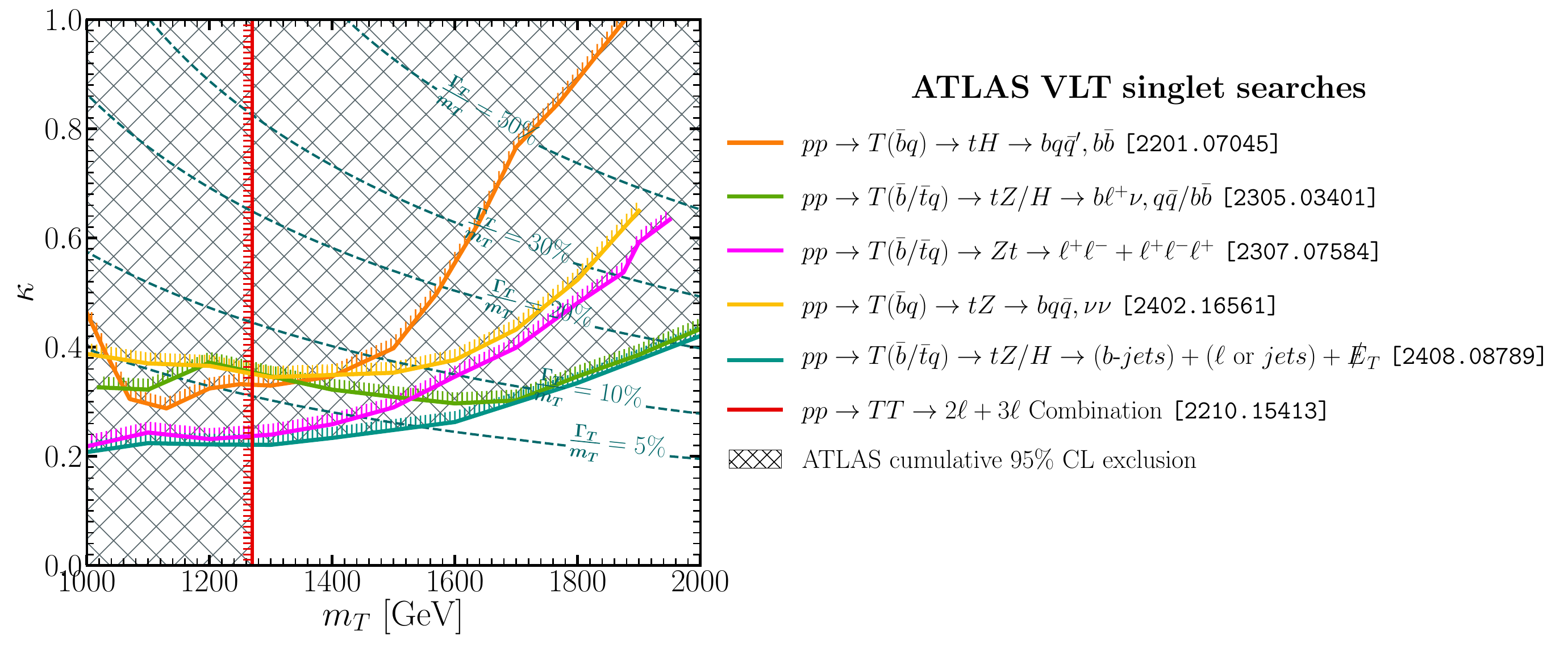}
	\caption{95\% CL exclusion region (hatched area) for the singlet VLT scenario based on ATLAS searches. Solid lines indicate individual exclusion limits from single and pair production analyses.}
	\label{fig:2}
\end{figure}

\begin{figure}[H]
	\centering
	\includegraphics[height=7cm,width=16cm]{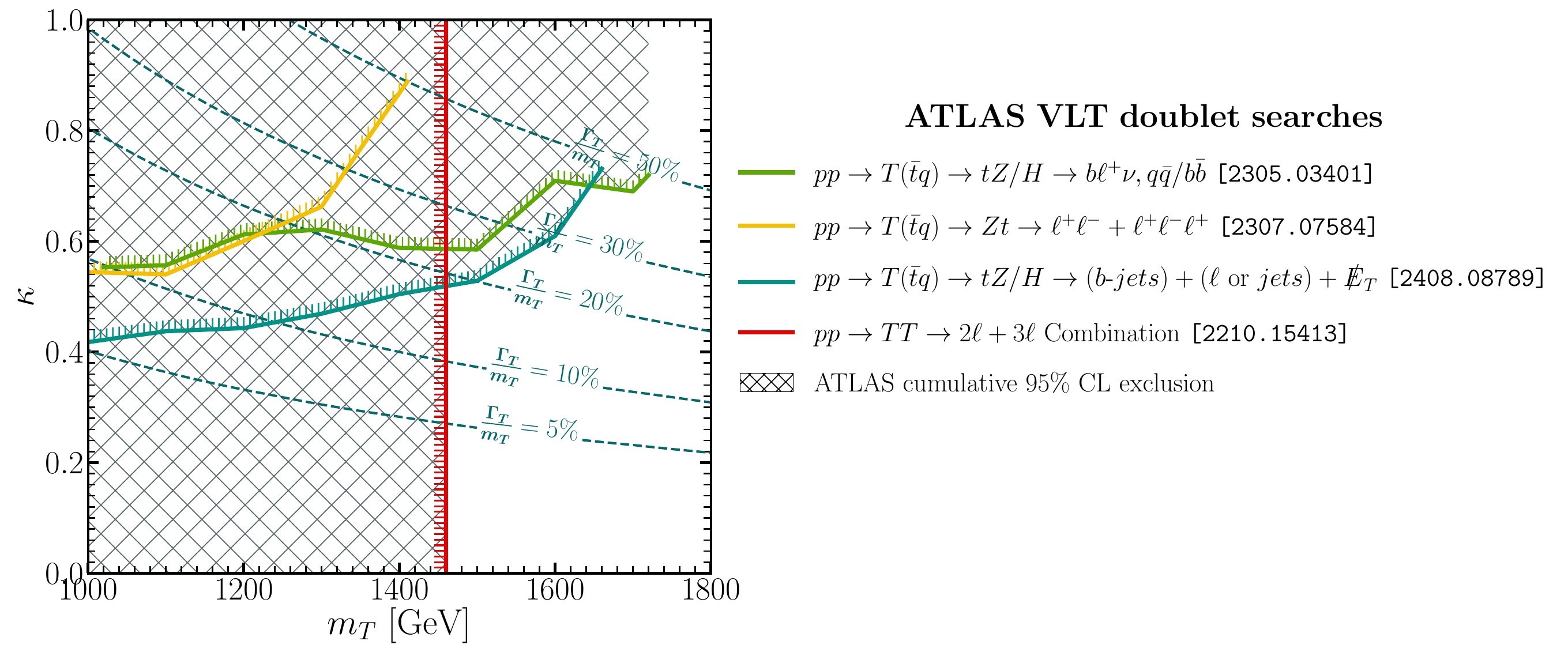}
	\caption{95\% CL exclusion limits for the doublet VLT scenario from ATLAS searches, analogous to Fig.~\ref{fig:2}.}
	\label{fig:3}
\end{figure}

Fig.~\ref{fig:2} presents the ATLAS 95\% CL exclusion limits for the singlet $T$ configuration in the $(m_T, \kappa)$ plane. Hatched regions represent the excluded parameter space, while solid lines indicate individual limits derived from both single and pair production analyses. The decay channels analyzed include $pp \rightarrow T \rightarrow tZ/h$, leading to final states such as $b\ell^+ \nu$, $q\bar{q}/b\bar{b}$ \cite{ATLAS:2022ozf}, $2\ell + 3\ell$ \cite{ATLAS:2023bfh}, and $(b\text{-jets}) + (\ell \text{ or jets}) + \cancel{E}_T$ \cite{ATLAS:2024xne}. For masses in the range $m_T \in [1, 1.4]$ TeV, the exclusion limits reach $\kappa > 0.21$, extending to $\kappa > 0.4$ near $m_T = 2$ TeV. These constraints, based on $\sqrt{s} = 13$ TeV data with an integrated luminosity of 139 fb$^{-1}$, are among the most stringent VLQ limits reported by ATLAS. Notably, the pair production analysis from Ref.~\cite{ATLAS:2022hnn}, utilizing $2\ell$ and $3\ell$ final states, provides the strongest exclusions for a singlet $T$, ruling out masses below $m_T = 1.27$ TeV, as highlighted by the red line.

The variations in exclusion limits across different analyses are primarily driven by the choice of final states. For instance, final states involving multiple leptons, such as $2\ell$ and $3\ell$, benefit from reduced SM background contributions, enhancing their sensitivity compared to more background-dominated channels involving jets or $b$-jets. A detailed classification of the final-state multiplicities used in these searches is provided in Table~\ref{tab1}.

Fig.~\ref{fig:3} shows the exclusion limits for the doublet $T$ configuration, which follow a similar structure to the singlet scenario depicted in Fig.~\ref{fig:2}. Single production analyses target the channels $pp \rightarrow T \rightarrow tZ/H$, yielding final states comparable to those in the singlet configuration \cite{ATLAS:2022ozf, ATLAS:2023bfh, ATLAS:2024xne}. These searches constrain $\kappa$ values to be below 0.5 for $m_T$ in the range $[1, 1.5]$ TeV, with the limits relaxing to $\kappa \leq 0.7$ for $m_T > 1.6$ TeV. The strongest single production constraints arise from Ref.~\cite{ATLAS:2024xne}. In pair production, the analysis from Ref.~\cite{ATLAS:2022hnn}, based on $2\ell$ and $3\ell$ final states, sets the most stringent exclusion, ruling out masses below 1.46 TeV.

The difference in exclusion limits between the singlet and doublet configurations is primarily attributed to the branching fractions of $T$, which vary depending on its representation. For singlet $T$, the branching fractions are typically $\mathcal{BR}(T \to Wb) : \mathcal{BR}(T \to Zt) : \mathcal{BR}(T \to ht) = 1/2 : 1/4 : 1/4$. In contrast, for doublet $T$, they are $\mathcal{BR}(T \to Wb) : \mathcal{BR}(T \to Zt) : \mathcal{BR}(T \to ht) = 0 : 1/2 : 1/2$. This difference directly affects the signal production rates in specific channels, influencing the sensitivity of the corresponding analyses.

\begin{figure}[H] 
	\centering \includegraphics[height=7.2cm,width=16cm]{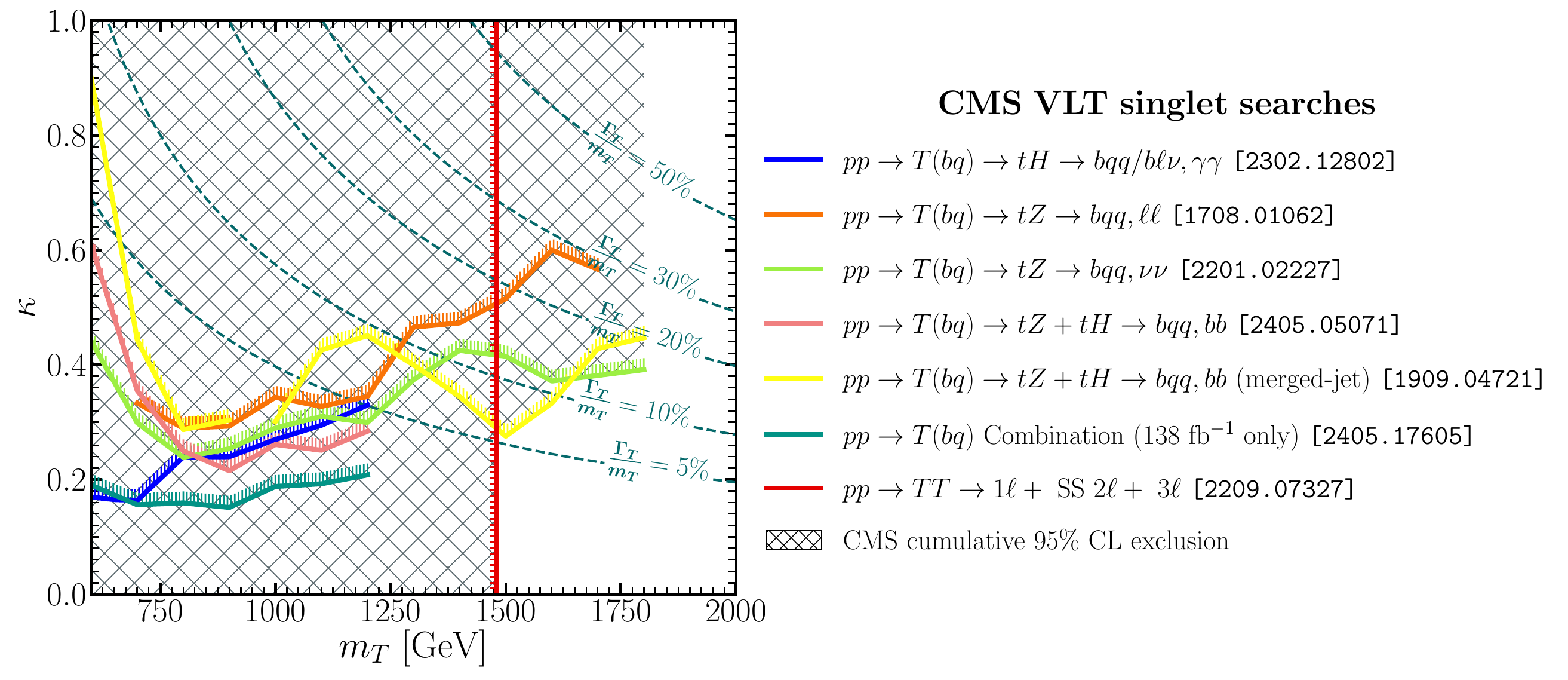} \caption{95\% CL exclusion region (hatched area) for the singlet VLT scenario based on CMS searches. Solid lines indicate individual exclusion limits from single and pair production analyses.} \label{fig:4}
\end{figure}
\begin{figure}[H]
	\centering \includegraphics[height=7.2cm,width=16cm]{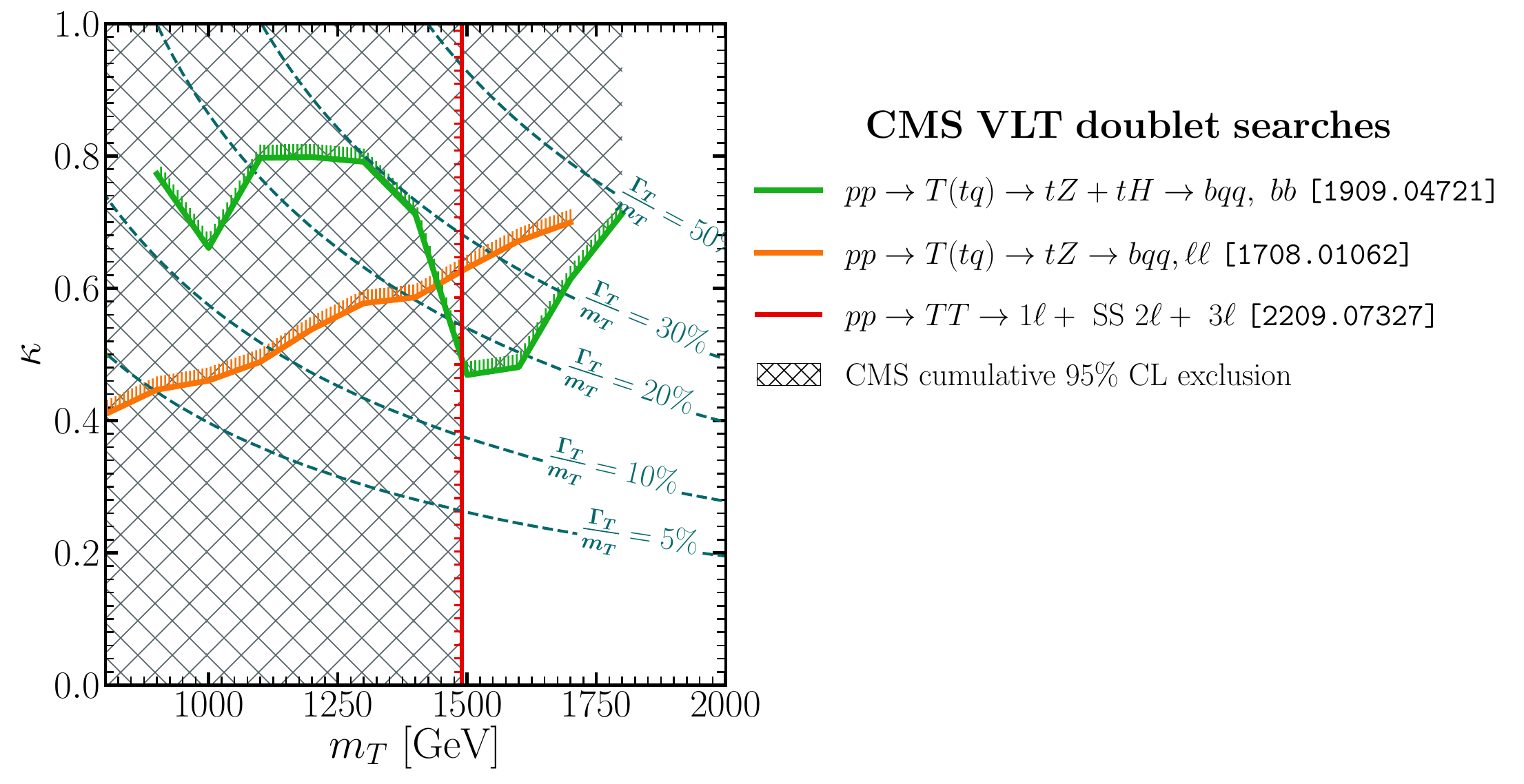} \caption{Similar to Fig.~\ref{fig:4}, but for the doublet VLT scenario.} \label{fig:5}
\end{figure}

CMS data on singlet and doublet $T$ quark searches \cite{CMS:2023agg,CMS:2017voh,CMS:2022yxp,CMS:2024qdd,CMS:2019afi,CMS:2024bni}, displayed in Figs.~\ref{fig:4} and \ref{fig:5}, provide refined exclusion limits. For the singlet $T$ configuration, the exclusion region begins at $m_T > 1.48$ TeV for pair production, derived from analyses covering final states with $1\ell$, $\text{SS}~2\ell$, and $3\ell$ signatures \cite{CMS:2022fck}, establishing the most stringent limits among all CMS searches for a singlet $T$. In single production, the relevant channels include $pp \rightarrow T (bq) \rightarrow th$ and $tZ$, with final states such as $bqq/b\ell\nu, \gamma\gamma$ \cite{CMS:2023agg}, $bqq, \ell\ell$ \cite{CMS:2017voh}, $bqq, \nu\nu$ \cite{CMS:2022yxp}, $bqq, bb$ \cite{CMS:2024qdd}, and $bqq, bb$ (merged-jet) \cite{CMS:2019afi}. The strongest limit arises from the combined analysis of these final states with 138 fb$^{-1}$ \cite{CMS:2024bni}, which excludes $\kappa \geq 0.16$ for $m_T \in [1, 1.2]$ TeV (a region already excluded by pair production), gradually relaxing to $\kappa = 0.3$ at $m_T = 1.5$ TeV and extending to $\kappa = 0.4$ at $m_T = 1.8$ TeV. In the region $m_T \geq 1.5$ TeV, the most stringent limits are provided by \cite{CMS:2019afi,CMS:2022yxp}.

For the doublet $T$ configuration, shown in Fig.~\ref{fig:5}, CMS results indicate slightly less stringent exclusion limits. The allowed region begins at $m_T > 1.49$ TeV for pair production, analyzed with final states covering $1\ell$, $\text{SS}~2\ell$, and $3\ell$ combinations \cite{CMS:2022fck}. For single production, decay channels include $pp \rightarrow T (tq) \rightarrow tZ + tH$, yielding final states such as $bqq, bb$ and $bqq, \ell\ell$ \cite{CMS:2024qdd}. The exclusion limit for the doublet configuration permits $\kappa$ values up to 0.4 for $m_T < 1.8$ TeV, broadening to 0.7 as the mass approaches 1.8 TeV.

The observed differences in exclusion limits between ATLAS and CMS, and across singlet and doublet configurations, stem from the same factors previously discussed, including the choice of final states, integrated luminosity, and branching fraction assumptions. As noted earlier, these branching fractions differ between singlet and doublet configurations, directly influencing the exclusion limits. Additional details on final-state multiplicities are provided in Table~\ref{tab2}, offering further insight into these differences.

\subsection{VLB Searches}

Fig.~\ref{fig:6} presents the ATLAS exclusion limits for the singlet $B$ configuration, extending the excluded region to $m_B > 1.33$ TeV. This constraint represents the most stringent limit derived from ATLAS pair production searches, focusing on final states such as $\ell + \geq 4 \, \text{jets} \, (1b\text{-tagged}) + \slashed{E}_T$ \cite{ATLAS:2022tla}. For single production, ATLAS analyses of final states involving $bH \rightarrow b, b \bar{b}$ \cite{ATLAS:2023ixh} set additional constraints, limiting $\kappa$ values to below 0.42 at $m_B = 1.1$ TeV, with the upper bound relaxing to approximately 0.6 for masses beyond $m_B \geq 1.3$ TeV. These limits are derived from data collected at $\sqrt{s} = 13$ TeV with an integrated luminosity of 139 fb$^{-1}$.

\begin{figure}[H] 
	\centering \includegraphics[height=7.2cm,width=16.cm]{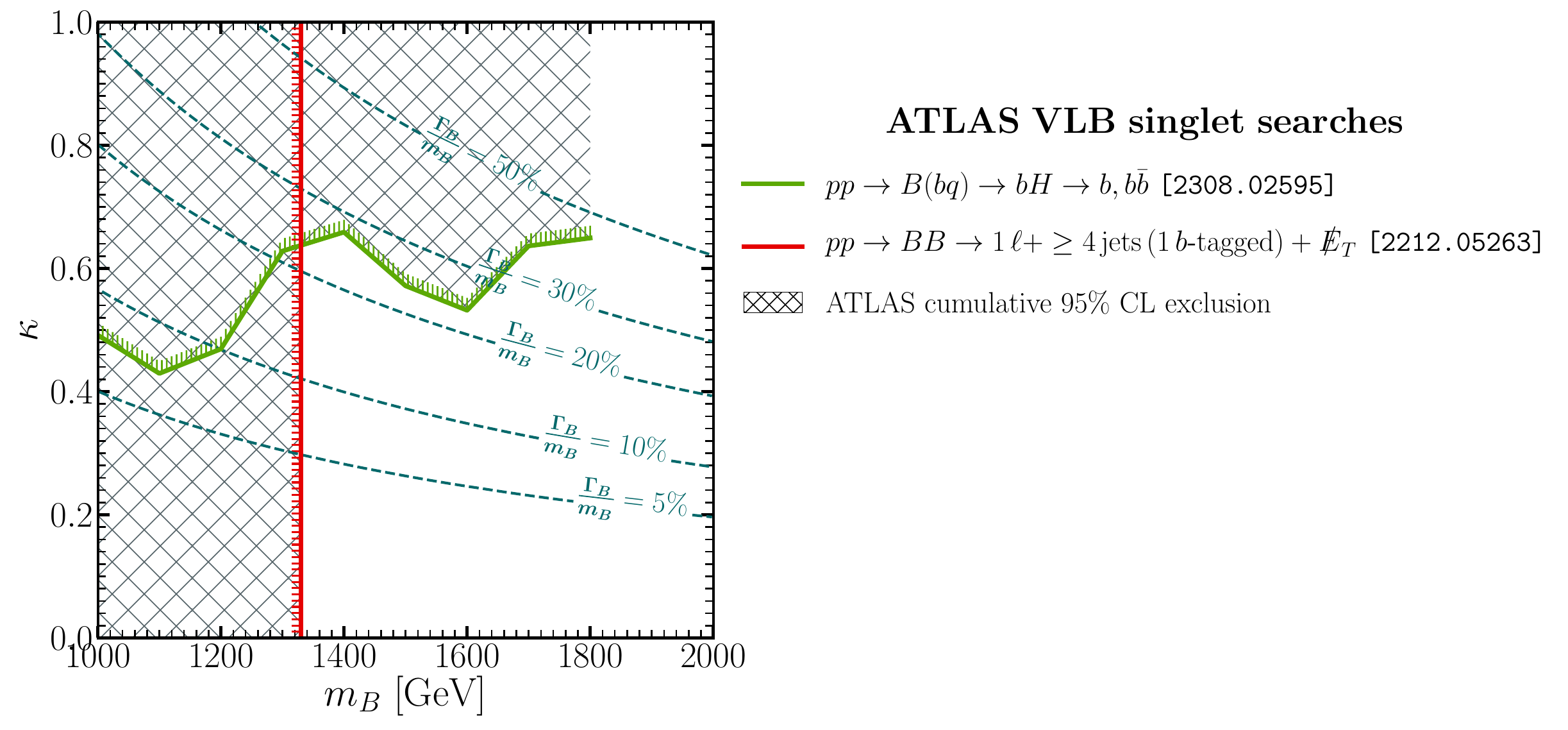} \caption{95\% CL excluded region (hatched area) for the singlet VLB scenario based on ATLAS searches. Solid lines indicate individual exclusion limits from single and pair production analyses.} \label{fig:6} 
\end{figure}

\begin{figure}[H]
	\centering \includegraphics[height=7.2cm,width=16.cm]{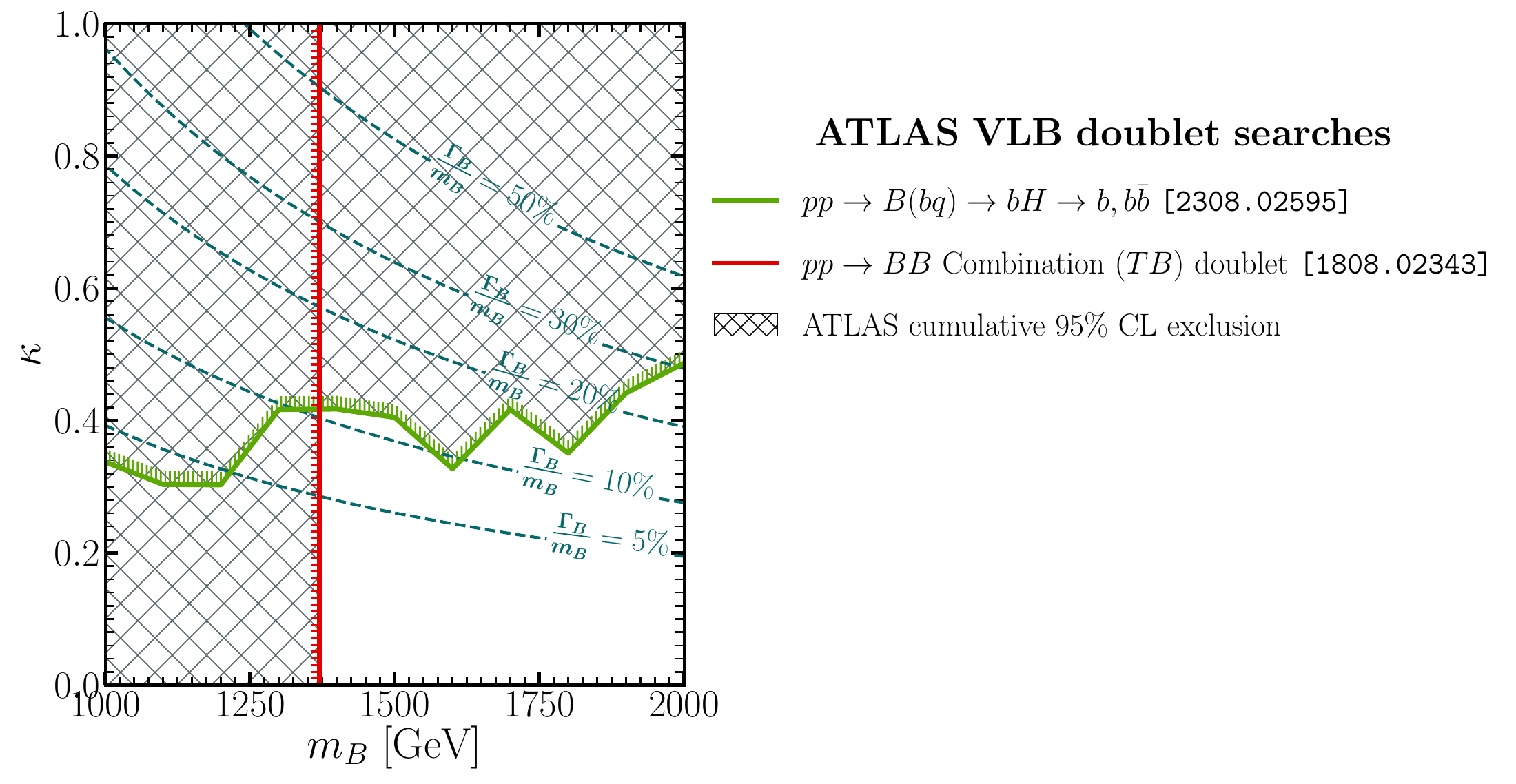} \caption{Similar to Fig.~\ref{fig:6}, but for doublet VLB.} \label{fig:7}
\end{figure}

For the doublet $B$ configuration, shown in Fig.~\ref{fig:7}, ATLAS extends exclusion limits to $m_B > 1.37$ TeV in pair production analyses. This limit, achieved using final states including $W(\ell\nu)t+X$, $Z(\ell\ell)t/b+X$, 3$\ell$, SS 2$\ell$, and fully hadronic modes \cite{ATLAS:2018ziw}, is the strongest exclusion reported by ATLAS for this configuration. Single production analyses using similar final states to the singlet configuration \cite{ATLAS:2023ixh} further restrict $\kappa$ to values below 0.3 for masses up to 1.2 TeV, with limits relaxing to $\kappa \leq 0.46$ for masses near 2 TeV. Detailed information on the final state multiplicities used in these searches is available in Table~\ref{tab3}.
\begin{figure}[H]
	\centering
	\includegraphics[height=7.5cm,width=16cm]{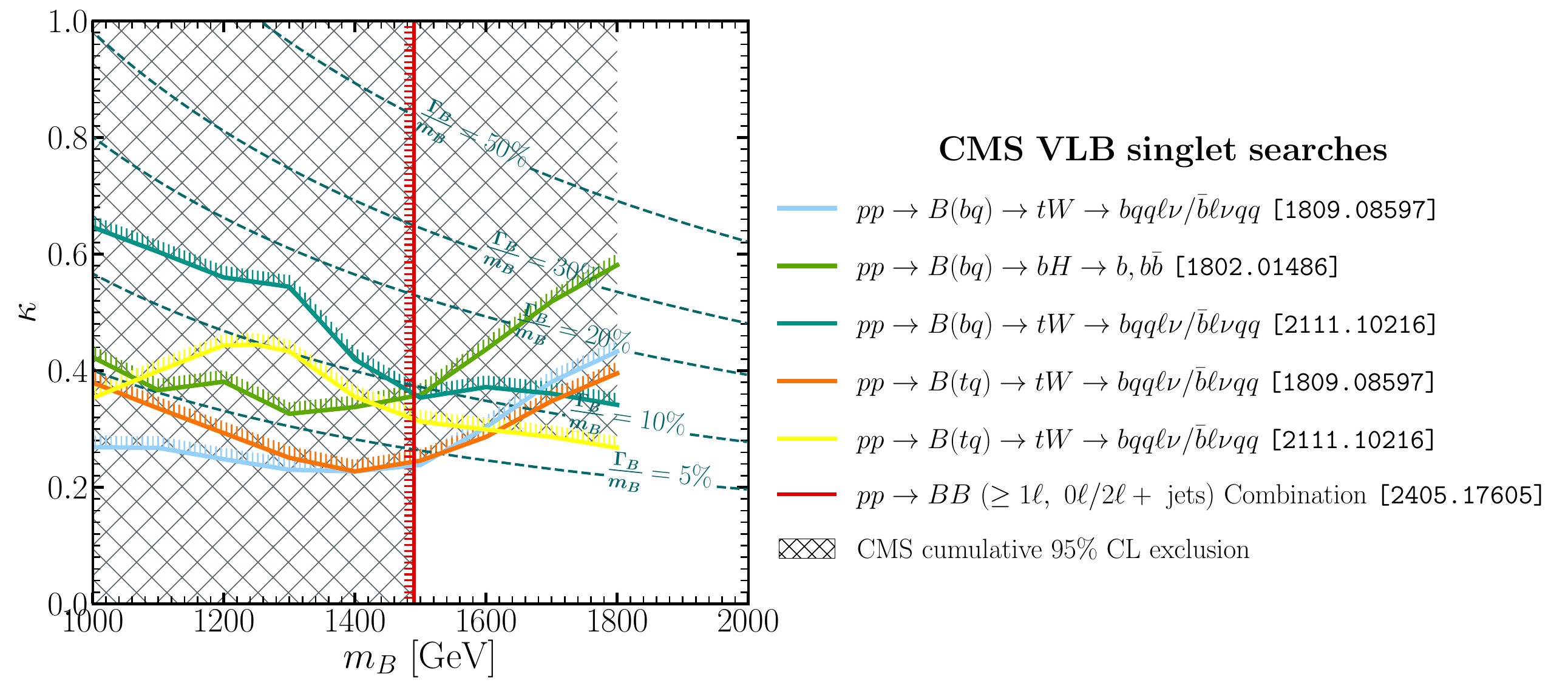}
	\caption{CMS exclusion limits for the singlet $B$ configuration, analogous to the ATLAS results shown in Fig.~\ref{fig:6}.}
	\label{fig:8}
\end{figure}

\begin{figure}[H]
	\centering
	\includegraphics[height=7.2cm,width=16cm]{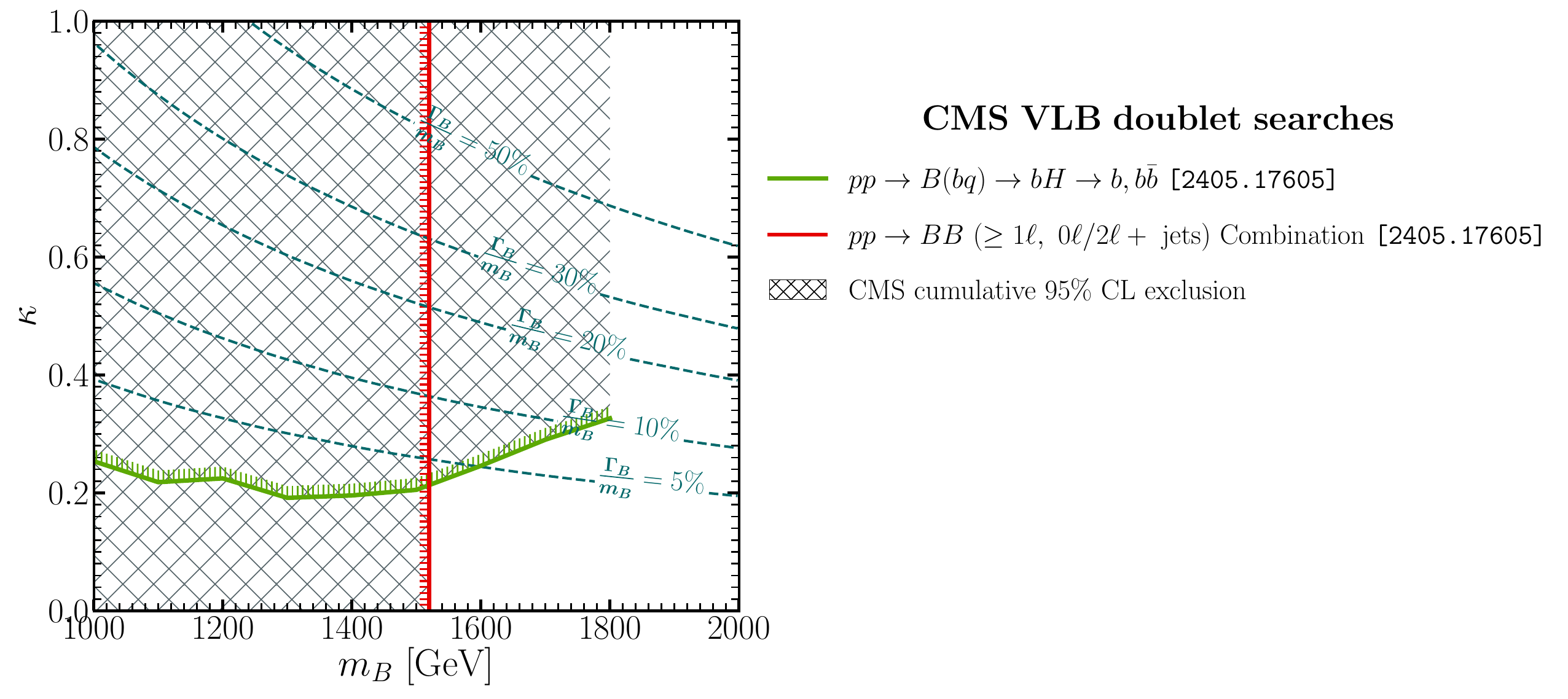}
	\caption{CMS exclusion limits for the doublet $B$ configuration, analogous to the ATLAS results shown in Fig.~\ref{fig:7}.}
	\label{fig:9}
\end{figure}
CMS exclusion limits for VLB searches, shown in Figs.~\ref{fig:8} and \ref{fig:9}, consistently provide the most stringent constraints for both singlet and doublet configurations in pair and single production modes. 

In the singlet $B$ configuration (Fig.~\ref{fig:8}), CMS pair production analyses exclude masses below 1.49 TeV, using final states such as $\geq 1\ell$, $0\ell/2\ell + \text{jets}$ \cite{CMS:2024bni}. For single production, analyses of channels like $bqql\nu/b\ell\nu qq$ \cite{CMS:2018dcw, CMS:2021mku} and $bH \rightarrow b, bb$ \cite{CMS:2018kcw} impose additional constraints. The strongest limits are provided by \cite{CMS:2018dcw}, which excludes $\kappa \geq 0.25$ for masses between 1 TeV and 1.5 TeV, and \cite{CMS:2021mku}, which excludes $\kappa \geq 0.26$ for masses ranging from 1.6 to 1.8 TeV.

For the doublet $B$ configuration (Fig.~\ref{fig:9}), CMS pair production analyses extend exclusions to $m_B > 1.52$ TeV, with limits based on final states such as $\geq 1\ell$, $0\ell/2\ell + \text{jets}$ \cite{CMS:2024bni}. In single production, the most restrictive limits arise from final states involving $bH \rightarrow b, bb$ \cite{CMS:2024bni}, with $\kappa$ constrained to values below 0.2 for masses up to 1.5 TeV, gradually relaxing to 0.32 for masses near 1.8 TeV. A detailed summary of the final state multiplicities used in CMS analyses is presented in Table~\ref{tab4}.

Consistent with the findings for VLT searches, the differences in exclusion limits between ATLAS and CMS, as well as those between singlet and doublet configurations, stem from variations in final state selections, integrated luminosity, and detector-specific sensitivities. Moreover, the branching fraction assumptions for $B$ quarks in the singlet and doublet representations play a significant role in shaping these limits. Notably, for VLB searches, CMS sets the most stringent constraints in both pair and single production modes, reflecting the high sensitivity achieved in these analyses.

\subsection{VLQ $X$ and $Y$ Searches}

This section examines the search strategies for the exotic VLQs $X$ and $Y$, for which fewer analyses exist compared to the VLT and VLB scenarios. Consequently, ATLAS and CMS exclusion limits are presented together in Figs.~\ref{fig:10} and \ref{fig:11} for $X$ and $Y$, respectively, covering results from both single and pair production modes. For additional details on the final state multiplicities used in these searches, see Tables~\ref{tab5} and \ref{tab6}.

In Fig.~\ref{fig:10}, which shows the exclusion limits for the $X$ VLQ, CMS data from pair production exclude masses below $m_X = 1.33$ TeV, using final states such as SS $2\ell$ and $1\ell$ channels combined with jets \cite{CMS:2018ubm}. ATLAS provides a more stringent exclusion, ruling out $m_X < 1.46$ TeV with final states involving $1\ell + \text{jets} + \slashed{E}_T$ \cite{ATLAS:2022tla}, marking the strongest LHC exclusion for $X$ VLQs in pair production.

For single production of the $X$ quark, several analyses by CMS have investigated channels like $pp \rightarrow X(tq) \rightarrow Wt$, producing a variety of final states. These include $bqq$, $\ell\nu/b\ell\nu$ and $qq$ \cite{CMS:2018dcw, CMS:2021mku}. In particular, CMS \cite{CMS:2018dcw} excludes $\kappa$ values greater than 0.16 for masses in the range $m_X \in [0.8, 1.6]$ TeV. A more restrictive limit is achieved in the higher mass region $m_X \in [1.6, 1.8]$ TeV, excluding $\kappa$ values above 0.2 \cite{CMS:2021mku}. Additionally, ATLAS explores both single and pair production with SS $2\ell$ and $3\ell$ final states accompanied by $b$-jets \cite{ATLAS:2018alq}. Notably, these exclusions also apply to $B$ quarks when assuming a pure decay scenario, i.e., $\mathcal{BR}(B \rightarrow Wt) = 100\%$ (for the corresponding ${\cal BR}$).
\begin{figure}[H]
	\centering
	\includegraphics[height=7.2cm,width=16cm]{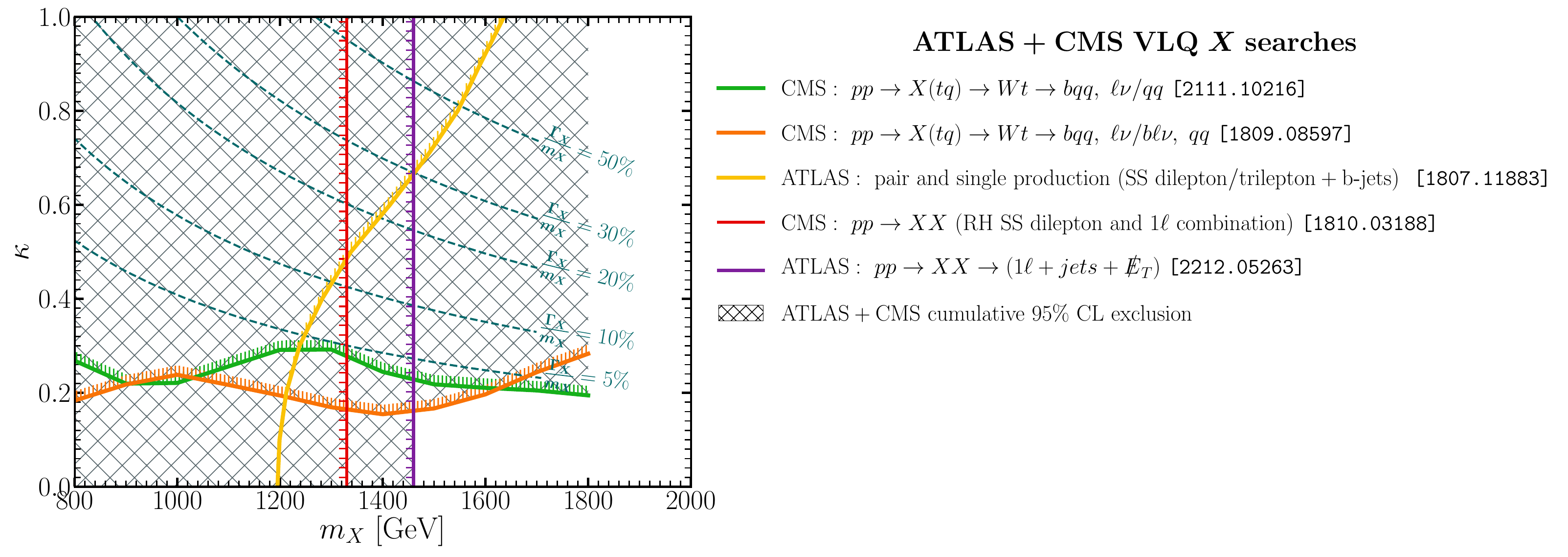}
	\caption{ATLAS and CMS exclusion limits for the $X$ VLQ.}
	\label{fig:10}
\end{figure}

\begin{figure}[H]
	\centering
	\includegraphics[height=7.2cm,width=16cm]{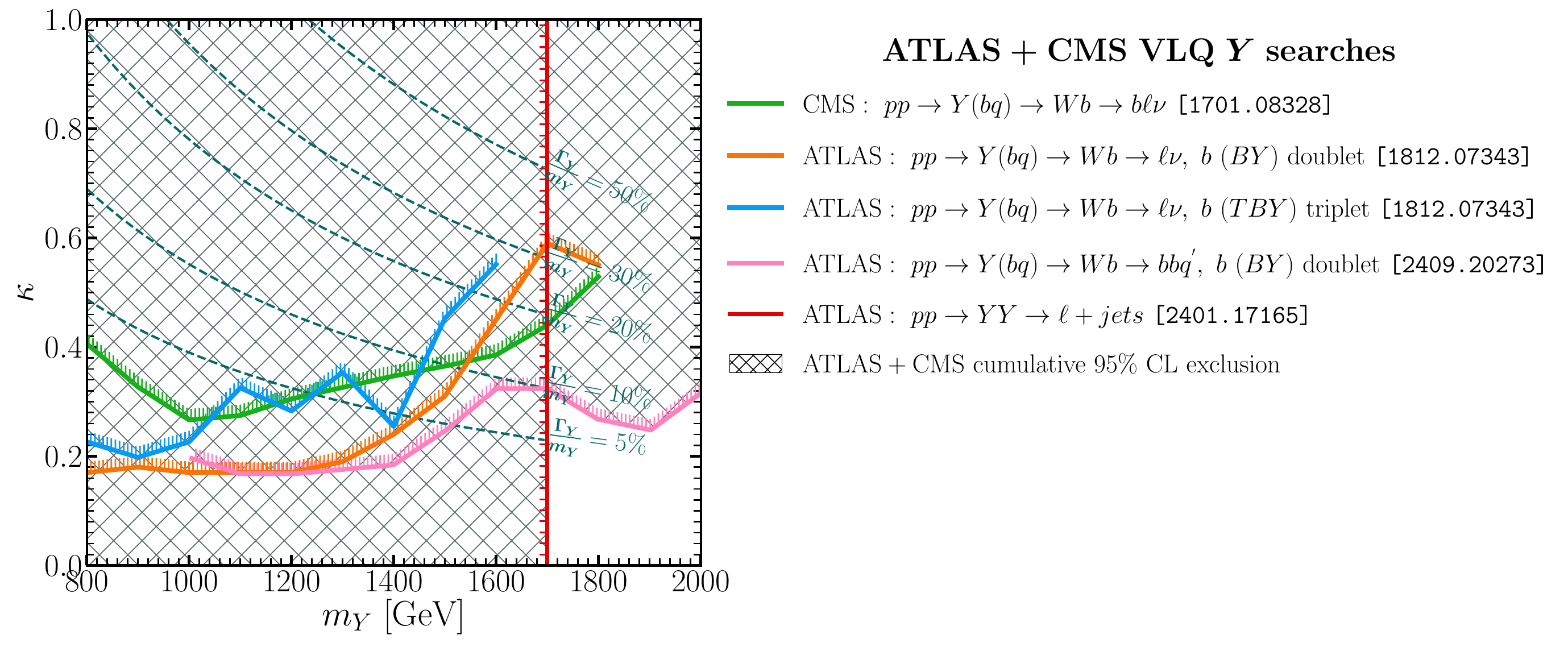}
	\caption{ATLAS and CMS exclusion limits for the $Y$ VLQ.}
	\label{fig:11}
\end{figure}
In Fig.~\ref{fig:11}, depicting results for the $Y$ VLQ, ATLAS provides the most stringent exclusion limits, ruling out masses below $m_Y = 1.7$ TeV in pair production with final states involving leptons and jets \cite{ATLAS:2024gyc}. This exclusion limit also applies to $T$ quarks in cases where the $T \rightarrow Wb$ decay occurs with a branching ratio of 100\%.

In single production of the $Y$ quark, CMS investigates the $b\ell\nu$ final state \cite{CMS:2017fpk}, while ATLAS examines multiple final states within the doublet $BY$ scenario, including $b\ell\nu$ \cite{ATLAS:2018dyh} and $bbbq'$ \cite{ATLAS:2024kgp}. The strongest exclusion limit for single production is set by ATLAS in the $bbbq'$ final state, where $\kappa \geq 0.3$ is excluded for $m_Y$ near 2 TeV \cite{ATLAS:2024kgp}. For lower masses, ATLAS excludes $\kappa \geq 0.17$ in the $b\ell\nu$ final state for $m_Y$ in the range [0.8, 1.2] TeV \cite{ATLAS:2018dyh}.

The exclusion limits for $X$ and $Y$ reflect their single dominant decay channels ($X \to Wt$ and $Y \to Wb$), resulting in similar final state structures featuring $W$ bosons and $b$-quarks. However, the presence of an intermediate top quark in $X$ decays introduces additional complexity, such as boosted $b$-jets and higher multiplicities in jets and leptons, enhancing the sensitivity of some analyses. Differences in exclusion limits between ATLAS and CMS are influenced by integrated luminosity, search strategies, and detector performance.

\begin{table}[H]
	\centering
	\def\arraystretch{1.1} 
	\resizebox{\textwidth}{!}{%
		\begin{tabular}{c|c|l|c|c|c|c|c}
			\toprule
			\textbf{VLQ} & \textbf{Decay Mode} & \textbf{Experiment} & \boldmath$\sqrt{s}$ & \textbf{Dataset} & \multicolumn{3}{c}{\textbf{Final State Multiplicities}} \\
			\cmidrule{6-8}
			& & & \textbf{[TeV]} & & $N_\ell$ & $N_b$ & $N_j$ \\
			\toprule
			\multirow{6}{*}
			& $Zt$ & ATLAS ~\cite{ATLAS:2017vdo} & 13 TeV & $36.1$ fb$^{-1}$  & $1\ell$	&$\geq 1$&	$\geq 4$ \\
			&  $Wb$ & ATLAS ~\cite{ATLAS:2017nap} & 13 TeV & $36.1$ fb$^{-1}$  & $1\ell$&	$\geq 1$&	$\geq 4$\\
			& $Wb$  & ATLAS ~\cite{ATLAS:2018alq} &  13 TeV & $36.1$ fb$^{-1}$  & $2\ell$ (SS), $3\ell$	&$\geq 1$&	$\geq 2$\\
			& $Wb$ & ATLAS ~\cite{ATLAS:2018dyh} &  13 TeV & $36.1$ fb$^{-1}$  & $1\ell$&	$\geq 1$&	$\geq 2$\\
			& $Ht$ & ATLAS ~\cite{ATLAS:2022ozf} & 13 TeV & $139$ fb$^{-1}$  & $0\ell$&	$\geq 2$&	$\geq 2$\\
			& Inclusive & ATLAS ~\cite{ATLAS:2022tla} &  13 TeV & $139$ fb$^{-1}$  & $1\ell$&	$\geq 1$&	$\geq 4$\\
			
			& $Ht, Zt$ & ATLAS ~\cite{ATLAS:2018cye}&  13 TeV & $36.1$ fb$^{-1}$  & $1\ell$&	$\geq 3$&	$\geq 6$\\
			& $Zt$ & ATLAS ~\cite{ATLAS:2022hnn} & 13 TeV & $139$ fb$^{-1}$  & $2\ell$ (OS), $\geq 3\ell$&	$\geq 1$&	$\geq 4$\\
			& $Ht, Zt$  & ATLAS ~\cite{ATLAS:2023pja} & 13 TeV & $139$ fb$^{-1}$ & $1\ell$&	$\geq 1$&	$\geq 4$\\
			& $Zt$ & ATLAS ~\cite{ATLAS:2023bfh} & 13 TeV & $139$ fb$^{-1}$  &$2\ell$ (OS), $\geq 3\ell$&	$\geq 1$&	$\geq 4$\\

			& $Zt$ & ATLAS ~\cite{ATLAS:2024xne} & 13 TeV & $139$ fb$^{-1}$  & $0\ell$&	$\geq 1$&	$\geq 2$\\
			$T$  & Inclusive & ATLAS ~\cite{ATLAS:2024gyc} & 13 TeV & $140$ fb$^{-1}$  & $1\ell$ & $\geq 1$&	$\geq 4$ \\
			& $Zt$  & ATLAS ~\cite{ATLAS:2018cjd} & 13 TeV & $36.1$ fb$^{-1}$  & $0,~1\ell$&	$\geq 1$&	$\geq 3$, $\geq 4$  \\ 
			& $Wb$ & ATLAS ~\cite{ATLAS:2016ovj} & 13 TeV & $3.2$ fb$^{-1}$  & $1\ell$&	$\geq 1$&	$\geq 2$\\
			& Inclusive & ATLAS ~\cite{ATLAS:2016btu} & 13 TeV & $13.2$ fb$^{-1}$  &  $0\ell$, $1\ell$&	$\geq 3$&	$\geq 4$\\
			&  Inclusive  & ATLAS ~\cite{ATLAS:2016sno} & 13 TeV & $3.2$ fb$^{-1}$  &  $\geq 2\ell$ (SS)&	$\geq 1$&	$\geq 2$\\
			& $Wb$ & ATLAS ~\cite{ATLAS:2016scx} & 8 TeV & $20.3$ fb$^{-1}$  & $1\ell$&	$\geq 1$&	$\geq 2$\\
			& $Zt$ & ATLAS ~\cite{ATLAS:2014vpn} & 8 TeV & $20.3$ fb$^{-1}$  & $\geq 2\ell$ (OS)&	$\geq 1$&	$\geq 2$ \\
			& Inclusive & ATLAS ~\cite{ATLAS:2015ktd} & 8 TeV & $20.3$ fb$^{-1}$  & $1\ell$&	$\geq 1$&	$\geq 4$\\
			& Inclusive & ATLAS ~\cite{ATLAS:2015uaw} & 8 TeV & $20.3$ fb$^{-1}$  &$\geq 2\ell$ (SS)&	$\geq 1$&	$\geq 4$\\
			&  Inclusive & ATLAS ~\cite{ATLAS:2018ziw} & 13 TeV & $36.1$ fb$^{-1}$  &$\geq 2\ell$ (SS)&	$\geq 2$&	$\geq 4$\\
			& $Ht,~Zt$ & ATLAS ~\cite{ATLAS:2024xdc} & 13 TeV &$139$ fb$^{-1}$  & $1\ell$, $\geq 2\ell$&	$\geq 2$&	$\geq 4$\\
			&  $Zt$ & ATLAS ~\cite{ATLAS:2018tnt} & 13 TeV & $36.1$ fb$^{-1}$  &$\geq 2\ell$, $\geq 3\ell$&	$\geq 1$&	$\geq 2$\\
			& $Wb$ & ATLAS ~\cite{ATLAS:2024kgp}&  13 TeV & $139$ fb$^{-1}$  & $0\ell$&	$\geq 1$&	$\geq 2$\\
			
			& $Ht$ & ATLAS ~\cite{ATLAS:2018uky} &  13 TeV & $36.1$ fb$^{-1}$  & $0\ell$	&$\geq 2$&	$\geq 4$\\
			& $Zt$  & ATLAS ~\cite{ATLAS:2016seq} & 13 TeV & $3.2$ fb$^{-1}$  & $1\ell$&	$\geq 1$&	$\geq 4$\\
			& $Wb$ & ATLAS ~\cite{ATLAS:2012qe} & 7 TeV  & $4.7$ fb$^{-1}$  & $1\ell$ & $\geq 1$&	$\geq 3$ \\
			& $Wb$ & ATLAS ~\cite{ATLAS:2012tkh} & 7 TeV  & $1.04$ fb$^{-1}$  & $1\ell$ & $\geq 1$&	$\geq 3$\\
			\bottomrule
		\end{tabular}
	}
	\caption{Summary of ATLAS searches for single and pair production of VLT across Run 1 and Run 2. The table provides details on decay modes, datasets and final state multiplicities: number of leptons ($N_\ell$), $b$-tagged jets ($N_b$) and jets ($N_j$).}
	\label{tab1}
\end{table}

\begin{table}[H]
	\centering
	\def\arraystretch{1.2} 
	\resizebox{\textwidth}{!}{%
		\begin{tabular}{c|c|l|c|c|c|c|c}
			\toprule
			\textbf{VLQ} & \textbf{Decay Mode} & \textbf{Experiment} & \boldmath$\sqrt{s}$ & \textbf{Dataset} & \multicolumn{3}{c}{\textbf{Final State Multiplicities}} \\
			\cmidrule{6-8}
			& & & \textbf{[TeV]} & & $N_\ell$ & $N_b$ & $N_j$ \\
			\toprule
			\multirow{2}{*}
			& $Zt$ & CMS ~\cite{CMS:2022yxp} & 13 TeV & $137$ fb$^{-1}$ & $0\ell$ & $\geq 1$&	$\geq 4$\\
			& $Ht$ & CMS ~\cite{CMS:2016jce} & 13 TeV & $2.3$ fb$^{-1}$  & $0\ell$ & $\geq 2$&	$\geq 2$ \\
			& $Ht$ & CMS ~\cite{CMS:2016edj} & 13 TeV & $2.3$ fb$^{-1}$  & $1\ell$ & $\geq 2$&	$\geq 4$\\
			&  Inclusive & CMS ~\cite{CMS:2017ked} & 13 TeV & $2.6$ fb$^{-1}$  & $1\ell$ &$\geq 1$&	$\geq 4$\\
			& $Ht$ & CMS ~\cite{CMS:2023agg} & 13 TeV & $138$ fb$^{-1}$  & $1\ell$ & $\geq 1$&	$\geq 4$\\
			& $Ht,Zt$ & CMS ~\cite{CMS:2024qdd} & 13 TeV & $138$ fb$^{-1}$  & $0\ell$ & $\geq 2$&	$\geq 6$\\

			& $Ht,Zt$ & CMS ~\cite{CMS:2019afi} & 13 TeV & $35.9$ fb$^{-1}$  & $0\ell$ & $\geq 2$&	$\geq 6$\\ 
			& $Zt$ & CMS ~\cite{CMS:2017voh} & 13 TeV & $35.9$ fb$^{-1}$  & $2\ell$  (OS)& $\geq 1$&	$\geq 3$\\
			& Inclusive & CMS ~\cite{CMS:2016ete} & 13 TeV & $2.3$ fb$^{-1}$  & $1\ell$ & $ \geq 1$ & $\geq 4$\\
			$T$  & Inclusive & CMS ~\cite{CMS:2013hwy} & 8 TeV & $19.5$ fb$^{-1}$  & $\geq 1\ell$ & $\geq 1$ &  $\geq 4$ \\
			& Inclusive & CMS ~\cite{CMS:2018zkf} & 13 TeV & $35.9$ fb$^{-1}$  &  $\geq 1\ell$,  $2\ell$ (SS), $ \geq 3\ell$ & $ \geq 1$ & $\geq 4$\\
			& Inclusive & CMS ~\cite{CMS:2022fck} & 13 TeV & $138$ fb$^{-1}$  & $\geq 1\ell$,  $2\ell$ (SS), $ \geq 3\ell$ & $ \geq 1$ & $\geq 4$\\
			&$Zt$ & CMS ~\cite{CMS:2017gsh} & 13 TeV & $2.3$ fb$^{-1}$  &$2\ell$ (OS)&	$\geq 1$&	$\geq 2$\\
			
			& $Zt$  & CMS ~\cite{CMS:2018wpl} & 13 TeV & $35.9$ fb$^{-1}$ & $2\ell$ (OS)&	$\geq 1$&	$\geq 2$\\
			
			& $Wb$ & CMS ~\cite{CMS:2017fpk} & 13 TeV & $2.3$ fb$^{-1}$  & $1\ell$ &$\geq 1$&	$\geq 2$\\
			& $Wb$ & CMS ~\cite{CMS:2017ynm} & 13 TeV & $35.8$ fb$^{-1}$  & $1\ell$ & $\geq 1$&	$\geq 4$\\
			& Inclusive & CMS ~\cite{CMS:2019eqb} & 13 TeV & $36.1$ fb$^{-1}$  & $0\ell$ & $\geq 2$ & $\geq 6$\\
			&Inclusive& CMS ~\cite{CMS:2015lzl} & 13 TeV & $19.7$ fb$^{-1}$  & $\geq 1\ell$ &$\geq 1$&	$\geq 4$  \\
			& $Wb$ & CMS ~\cite{CMS:2012ab} & 7 TeV & $5.0$ fb$^{-1}$  & $ 2\ell$ (OS)&	$\geq 2$&	$\geq 2$\\
			& $Ht$ & CMS ~\cite{CMS:2015jwh} & 8 TeV & $19.7$ fb$^{-1}$  & $0\ell$ & 	$\geq 2$&	$\geq 6$\\
			& $Wb$ & CMS ~\cite{CMS:2012mir} & 7 TeV & $5.0$ fb$^{-1}$  & $ 1\ell$ &	$\geq 1$&	$\geq 4$\\
			\bottomrule
		\end{tabular}
	}
	\caption{Summary of CMS searches for VLT. Similar to Table~\ref{tab1}.}
	\label{tab2}
\end{table}
\begin{table}[H]
	\centering
	\def\arraystretch{1.2} 
	\resizebox{\textwidth}{!}{%
		\begin{tabular}{c|c|l|c|c|c|c|c}
			\toprule
			\textbf{VLQ} & \textbf{Decay Mode} & \textbf{Experiment} & \boldmath$\sqrt{s}$ & \textbf{Dataset} & \multicolumn{3}{c}{\textbf{Final State Multiplicities}} \\
			\cmidrule{6-8}
			& & & \textbf{[TeV]} & & $N_\ell$ & $N_b$ & $N_j$ \\
			\toprule
			\multirow{6}{*}
			&  $Wt$ & ATLAS ~\cite{ATLAS:2017nap} & 13 TeV & $36.1$ fb$^{-1}$  & $1\ell$ &$\geq 2$&	$\geq 4$\\
			& Inclusive & ATLAS ~\cite{ATLAS:2018alq} & 13 TeV & $36.1$ fb$^{-1}$  & $\geq 2\ell$ (SS)&	$\geq 1$ & $\geq 4$\\ 
			& Inclusive & ATLAS ~\cite{ATLAS:2022tla} &  13 TeV & $139$ fb$^{-1}$  & $1\ell$ & $\geq 1$	&$\geq 4$\\
			& $Zb$ & ATLAS ~\cite{ATLAS:2022hnn} & 13 TeV & $139$ fb$^{-1}$  & $\geq 2\ell$ (OS) or $\geq 3$ &	$\geq 1$	&$\geq 4$\\
			&  Inclusive & ATLAS ~\cite{ATLAS:2016sno} & 13 TeV & $3.2$ fb$^{-1}$  &  $\geq 2\ell$ (SS)&	$\geq 1$&	$\geq 4$\\
			 $B$  &  Inclusive & ATLAS ~\cite{ATLAS:2018ziw} & 13 TeV & $36.1$ fb$^{-1}$  & $\geq 1\ell$ &	$\geq 1$&	$\geq 4$\\
			&  $Zb$ & ATLAS ~\cite{ATLAS:2014vpn} & 8 TeV & $20.3$ fb$^{-1}$  & $\geq 2\ell$ (OS)&	$\geq 1$&	$\geq 2$\\
			& Inclusive & ATLAS ~\cite{ATLAS:2015ktd} & 8 TeV & $20.3$ fb$^{-1}$  & $\geq 1\ell$ & $\geq 2$ & $\geq 4$\\
			& Inclusive & ATLAS ~\cite{ATLAS:2015uaw} & 8 TeV & $20.3$ fb$^{-1}$  & $\geq 2\ell$ (SS)&	$\geq 1$&	$\geq 4$\\
			& $bH$ & ATLAS ~\cite{ATLAS:2023ixh} & 13 TeV &  $139$ fb$^{-1}$  & $0\ell$ &$\geq 1$&	$\geq 3$\\
			& $Wt$ & ATLAS ~\cite{ATLAS:2018mpo} & 13 TeV & $36.1$ fb$^{-1}$  & $1\ell$ & $\geq 1$&	$\geq 4$\\
			& Inclusive &  ATLAS ~\cite{ATLAS:2015vzd} & 8 TeV & $20.3$ fb$^{-1}$  & $1\ell$ & $\geq 1$ & $\geq 4$\\
			& Inclusive & ATLAS ~\cite{ATLAS:2018uky} &  13 TeV & $36.1$ fb$^{-1}$  & $0\ell$ &$ \geq 1$ & $\geq 4$\\
			&  $Zb$ & ATLAS ~\cite{ATLAS:2012wxk} & 7 TeV & $2$ fb$^{-1}$  & $2\ell$ (OS)&	$\geq 2$&	$\geq 4$\\
			\bottomrule
			
		\end{tabular}
	}
	\caption{Summary of ATLAS searches for VLB. This table follows the same format as Table~\ref{tab1}.}
	\label{tab3}
\end{table}
\begin{table}[H]
	\centering
	\def\arraystretch{1.2} 
	\resizebox{\textwidth}{!}{%
		\begin{tabular}{c|c|l|c|c|c|c|c}
			\toprule
			\textbf{VLQ} & \textbf{Decay Mode} & \textbf{Experiment} & \boldmath$\sqrt{s}$ & \textbf{Dataset} & \multicolumn{3}{c}{\textbf{Final State Multiplicities}} \\
			\cmidrule{6-8}
			& & & \textbf{[TeV]} & & $N_\ell$ & $N_b$ & $N_j$ \\
			\toprule
			\multirow{2}{*}
			&  Inclusive  & CMS ~\cite{CMS:2017ked} & 13 TeV & $2.6$ fb$^{-1}$  & $1\ell$ & $\geq1$   & $\geq4$ \\
			& Inclusive & CMS ~\cite{CMS:2018zkf} & 13 TeV & $35.9$ fb$^{-1}$  &  $1\ell$,  $2\ell$ (SS), $\geq3\ell$ & $\geq1$   & $\geq4 $\\
			& Inclusive & CMS ~\cite{CMS:2022fck} & 13 TeV & $138$ fb$^{-1}$  & $1\ell$,  $2\ell$ (SS), $\geq3\ell$ & $\geq1$   & $\geq4 $\\
			&$Zb$ & CMS ~\cite{CMS:2017gsh} & 13 TeV & $2.3$ fb$^{-1}$  &  $2\ell$ (OS) &  $\geq1$   & $\geq2 $\\
			& $Hb$ &  CMS ~\cite{CMS:2018kcw} & 13 TeV &  $35.9$ fb$^{-1}$ &  0$\ell$ &$\geq 1$&	$\geq 2$ \\
			& $Wt$ &  CMS ~\cite{CMS:2021mku} & 13 TeV & $138$ fb$^{-1}$ &  1$\ell$&  $\geq1$ &  $\geq2$   \\
			$B$& Inclusive &  CMS ~\cite{CMS:2024bni} & 13 TeV & $36$ fb$^{-1}$  & $0\ell$, $1\ell$, $2\ell$ (SS/OS), $\geq 3$  & $\geq1$ &  $\geq4$ \\
				& Inclusive &  CMS ~\cite{CMS:2024xbc} & 13 TeV & $138$ fb$^{-1}$  &  $0\ell$,  $2\ell$ (OS)& $\geq 2$&	$\geq 4$\\
			
			& $Zb$ &  CMS ~\cite{CMS:2018wpl} & 13 TeV & $35.9$ fb$^{-1}$  &  $2\ell$ (OS)&	$\geq 1$&	$\geq 2$ \\
			
			& $Hb, Zb$&  CMS ~\cite{CMS:2020ttz} & 13 TeV &  $138$ fb$^{-1}$  &  $0\ell$ & $\geq 2$&	$\geq 6$   \\
			
			& Inclusive &  CMS ~\cite{CMS:2019eqb} & 13 TeV & $36.1$ fb$^{-1}$  & $0\ell$ & $\geq 2$ & $\geq 6$ \\

			& Inclusive &  CMS ~\cite{CMS:2015hyy} & 8 TeV &  $19.7$ fb$^{-1}$ & $0\ell$, $1\ell$, $\geq 2\ell$ (OS/SS) & $\geq2$  & $\geq4$\\
			& $Wt$ &  CMS ~\cite{CMS:2018dcw} & 13 TeV & $35.9$ fb$^{-1}$  &  1$\ell$ & $\geq1$  & $\geq2$\\
			\bottomrule
		\end{tabular}
	}
	\caption{Summary of CMS searches for VLB. Similar to Table~\ref{tab3}.}
	\label{tab4}
\end{table}
\begin{table}[H]
	\centering
	\def\arraystretch{1.2} 
	\resizebox{\textwidth}{!}{%
		\begin{tabular}{c|c|l|c|c|c|c|c}
			\toprule
			\textbf{VLQ} & \textbf{Decay Mode} & \textbf{Experiment} & \boldmath$\sqrt{s}$ & \textbf{Dataset} & \multicolumn{3}{c}{\textbf{Final State Multiplicities}} \\
			\cmidrule{6-8}
			& & & \textbf{[TeV]} & & $N_\ell$ & $N_b$ & $N_j$ \\
			\toprule
			\multirow{6}{*}{$X$} 
			& $ Wt$ & ATLAS ~\cite{ATLAS:2017nap} & 13 & $36.1\,\text{fb}^{-1}$ & $1\ell$, $2\ell$ (OS) & $\geq 1$ & $\geq 4$ \\
			& $ Wt$ & ATLAS ~\cite{ATLAS:2018alq} & 13 & $36.1\,\text{fb}^{-1}$ & $2\ell$ (SS), $3\ell$& $\geq 1$ & $\geq 2$ \\
			& $Wt$ & ATLAS ~\cite{ATLAS:2018mpo} & 13 & $36.1\,\text{fb}^{-1}$ & $1\ell$ & $\geq 1$ & $\geq 4$ \\
			& $ Wt$ & ATLAS ~\cite{ATLAS:2016sno} & 13 & $3.2\,\text{fb}^{-1}$ & $2\ell$ (SS), $\geq 2\ell$ & $\geq 1$ & $\geq 2$ \\
			& $ Wt$ & ATLAS ~\cite{ATLAS:2015uaw} & 8 & $20.3\,\text{fb}^{-1}$ & $2\ell$ (SS), $\geq 2\ell$& $\geq 1$ & $\geq 2$ \\	& $ Wt$ & ATLAS ~\cite{ATLAS:2022tla} & 13 & $139\,\text{fb}^{-1}$ & $1\ell$ & $\geq 1$ & $\geq 4$ \\
			& $Wt$ & ATLAS ~\cite{ATLAS:2015vzd} & 8 & $20.3\,\text{fb}^{-1}$ & $1\ell$ & $\geq 1$ & $\geq 4$ \\
			
			\midrule
			\multirow{5}{*}{$Y$}
			& $Wb$ & ATLAS ~\cite{ATLAS:2016ovj} & 13 & $3.2\,\text{fb}^{-1}$ & $1\ell$ & $\geq 1$ & $\geq 2$ \\
			& $ Wb$ & ATLAS ~\cite{ATLAS:2017nap} & 13 & $36.1\,\text{fb}^{-1}$ & $1\ell$&	$\geq 1$&	$\geq 4$ \\
			& $ Wb$ & ATLAS ~\cite{ATLAS:2024gyc} & 13 & $140\,\text{fb}^{-1}$ & $1\ell$&	$\geq 1$&	$\geq 4$ \\
			& $ Wb$ & ATLAS ~\cite{ATLAS:2016scx} & 8 & $20.3\,\text{fb}^{-1}$ & $1\ell$&	$\geq 1$&	$\geq 2$ \\
			
			& $Wb$ & ATLAS ~\cite{ATLAS:2018dyh} & 13 & $36.1\,\text{fb}^{-1}$ & $1\ell$&	$\geq 1$&	$\geq 2$\\
			& $Wb$ & ATLAS ~\cite{ATLAS:2024kgp} & 13 & $139\,\text{fb}^{-1}$ & $0\ell$	&$\geq 1$&	$\geq 2$ \\
			\bottomrule
		\end{tabular}
	}
	\caption{Summary of ATLAS searches for VLQs $X$ and $Y$. This table follows the same format as Table~\ref{tab1}.}
	\label{tab5}
\end{table}

\begin{table}[H]
	\centering
	\def\arraystretch{1.2} 
	\resizebox{\textwidth}{!}{%
		\begin{tabular}{c|c|l|c|c|c|c|c}
			\toprule
			\textbf{VLQ} & \textbf{Decay Mode} & \textbf{Experiment} & \boldmath$\sqrt{s}$ & \textbf{Dataset} & \multicolumn{3}{c}{\textbf{Final State Multiplicities}} \\
			\cmidrule{6-8}
			& & & \textbf{[TeV]} & & $N_\ell$ & $N_b$ & $N_j$ \\
			\toprule
			\multirow{2}{*}{$X$} 
			& $Wt$ & CMS ~\cite{CMS:2018dcw} & 13 & $35.9\,\text{fb}^{-1}$ & $1\ell$&	$\geq 1$&	$\geq 2$\\
			& $Wt$ & CMS ~\cite{CMS:2018ubm} & 13 & $35.9\,\text{fb}^{-1}$ & $1\ell,~2\ell$ (SS)&	$\geq 1$&	$\geq 2$ \\
			&  $Wt$ & CMS ~\cite{CMS:2024xbc} & 13 & $138\,\text{fb}^{-1}$ &$0\ell,~2\ell$ (OS)&	$\geq 2$&	$\geq 4$ \\
			\midrule
			\multirow{2}{*}{$Y$} 
			& $Wb$ & CMS ~\cite{CMS:2017fpk} & 13 & $2.3\,\text{fb}^{-1}$ & $1\ell$&	$\geq 1$&	$\geq 2$ \\
			& $Wb$ & CMS ~\cite{CMS:2017ynm} & 13 & $35.8\,\text{fb}^{-1}$ & $1\ell$&	$\geq 1$&	$\geq 4$\\
			
			\bottomrule
		\end{tabular}
	}
	\caption{Summary of CMS searches for VLQs $X$ and $Y$. Similar to Table~\ref{tab5}.}
	\label{tab6}
\end{table}

\section{Global Exclusion from the LHC at 95\% CL}
\label{sec:Global}

This section presents a comparative analysis of the exclusion limits derived from both ATLAS and CMS data, identifying the most stringent exclusions across available datasets. By systematically selecting the strongest exclusion at each mass point, we delineate the 95\% CL exclusion regions for both single and pair production modes. 
\begin{figure}[H]
	\centering
	\includegraphics[height=19cm,width=15cm]{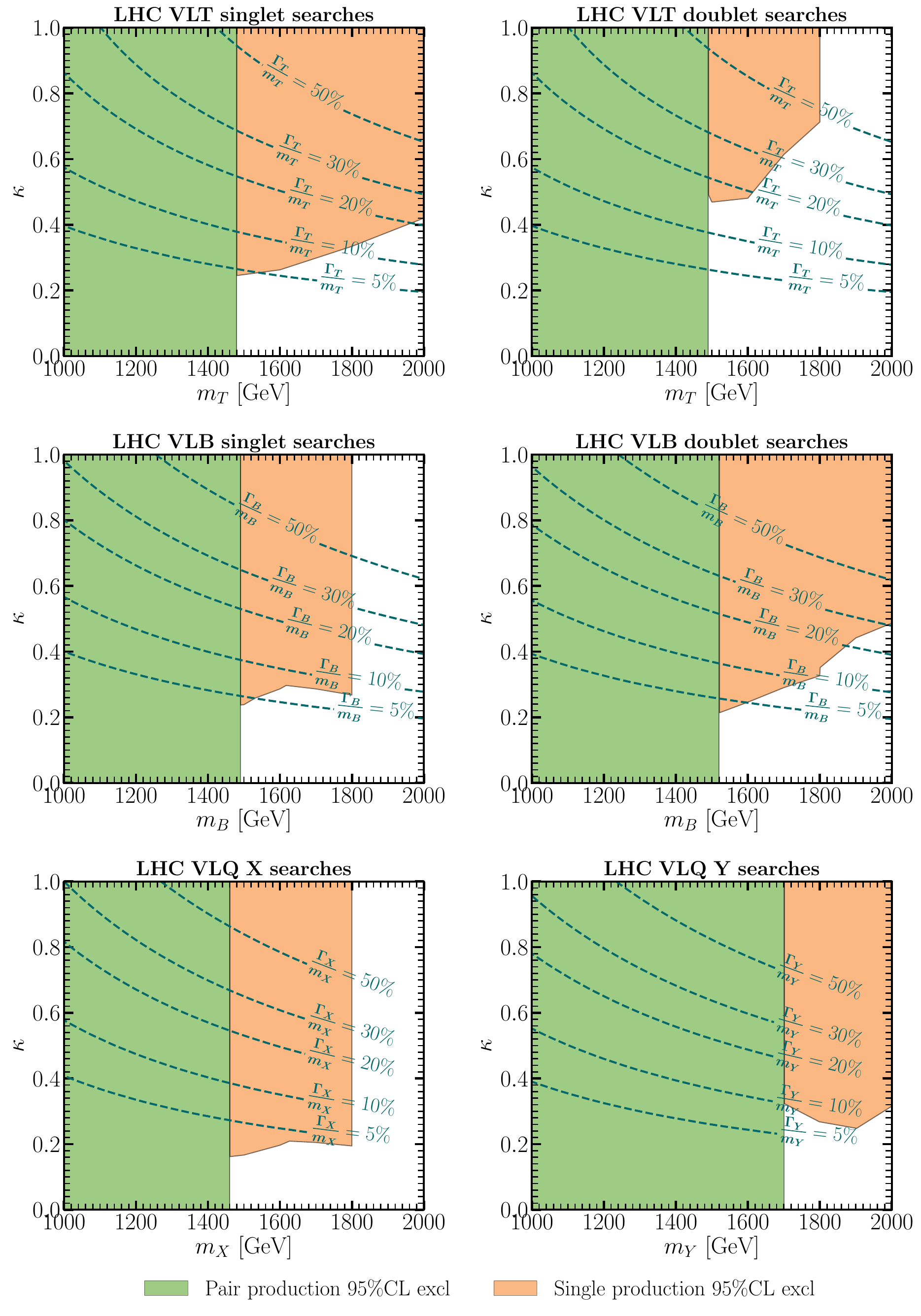}
	\caption{Joint LHC 95\% CL exclusion region for the singlet VLT (upper panels), VLB (middle) and VLQ $X$ and $Y$ (lower) scenarios, showing the most stringent limits from ATLAS and CMS results. Exclusion regions for single production are shown in orange, while pair production exclusions are shown in green.}
	\label{fig:12}
\end{figure}
Fig.~\ref{fig:12} summarizes these results: the upper panels display exclusion limits for the VLT singlet and doublet configurations, the middle panels present constraints for VLB and the lower panels illustrate exclusion regions for the exotic VLQs $X$ and $Y$. In these plots, orange regions represent exclusions from single production, while green regions denote limits from pair production. Detailed descriptions of the analyses and the associated final state multiplicities are provided in Tables\ref{tab1}--\ref{tab6}.

For the singlet VLT configuration, CMS achieves the strongest exclusion limit in pair production, ruling out masses below 1.48 TeV, based on final states that include $1\ell$, SS $2\ell$ and $3\ell$ channels \cite{CMS:2022fck}. In single production, ATLAS provides the most stringent constraints, restricting $\kappa$ to values below 0.26 around $m_T \sim 1.5$ TeV, with exclusions relaxing at higher masses, allowing $\kappa$ up to approximately 0.42 near $m_T = 2$ TeV \cite{ATLAS:2024xne}.

In the doublet VLT configuration, CMS again sets the strongest exclusion in pair production, excluding masses below $m_T = 1.49$ TeV based on final states involving $1\ell$, SS $2\ell$ and $3\ell$ channels \cite{CMS:2022fck}. For single production, CMS data yield the tightest constraints, restricting $\kappa$ to values below 0.46 at $m_T \sim 1.5$--$1.6$ TeV and expanding to approximately $\kappa = 0.7$ as the mass approaches $m_T = 1.8$ TeV, using final states such as $bqq$, $bb$ and $bqq, \ell\ell$ \cite{CMS:2024qdd}.

For the singlet VLB configuration, CMS provides the most stringent exclusion in pair production, ruling out masses below 1.49 TeV with a combination of final states including $\geq 1\ell$, $0\ell/2\ell + \text{jets}$ \cite{CMS:2024bni}. For single production, CMS also sets the strongest limits, with final states like $bqql\nu/b\ell\nu qq$ \cite{CMS:2018dcw, CMS:2021mku}, excluding $\kappa \geq 0.26$ for masses in the range 1.5 to 1.8 TeV.

In the doublet VLB configuration, CMS achieves the most restrictive exclusion limit in pair production, excluding masses below 1.52 TeV using final states such as $\geq 1\ell$, $0\ell/2\ell + \text{jets}$ \cite{CMS:2024bni}. For single production, CMS data offer the strongest exclusion limits with final states involving $bH \rightarrow b, bb$ \cite{CMS:2024bni}, constraining $\kappa$ to values below 0.2 for masses up to 1.52 TeV, gradually relaxing to $\kappa \leq 0.32$ as masses approach 1.8 TeV.

For the exotic VLQ $X$, ATLAS sets the most stringent exclusion, ruling out $m_X < 1.46$ TeV with final states involving $1\ell + \text{jets} + \slashed{E}_T$ \cite{ATLAS:2022tla}. For single production, CMS imposes the strongest constraints on $\kappa$, excluding values above 0.16 for masses in the range $m_X \in [1.46, 1.6]$ TeV \cite{CMS:2018dcw} and further tightening to exclude $\kappa$ values above 0.2 for masses in the region $m_X \in [1.6, 1.8]$ TeV \cite{CMS:2021mku}.

For the $Y$ VLQ, ATLAS provides the most stringent exclusion in pair production, excluding masses below 1.7 TeV using final states involving leptons and jets \cite{ATLAS:2024gyc}. In single production, ATLAS imposes the strongest constraints from the $BY$ doublet configuration, utilizing final states of $bbbq'$ \cite{ATLAS:2024kgp} to exclude $\kappa$ values above 0.26 for masses permitted by the pair production exclusion ($m_Y > 1.7$ TeV). 

The global exclusion regions reflect the interplay between production modes and VLQ representations, with differences primarily driven by the distinct $\mathcal{BR}$ assumptions associated to each configuration. Combined with variations in search strategies, integrated luminosity, and detector performance, these factors underscore the importance of systematically identifying the most stringent limits from available analyses to provide a comprehensive constraint on the VLQ parameter space.

 \section{Conclusion} \label{sec:conclusion}
This study provides a detailed analysis of exclusion limits on VLQs, drawing on extensive ATLAS and CMS data from the LHC. By examining single and pair production channels, we have identified the most stringent constraints on VLQ masses and mixing parameters across multiple final states, offering an integrated perspective on the current phenomenological landscape for VLQs at the CERN machine. In doing so, we have gone beyond standard practices by exploring bounds on doublets and triplets of VLQs in addition to the conventional paradigm of singlet VLQs. Furthermore, we did not assume degenerate masses for VLQs in doublet and triplet representations, as is often customary. Finally, we considered both single and pair production processes, highlighting their complementary roles in constraining VLQ scenarios.

To derive exclusion bounds, we adopted the most stringent experimental measurements provided by ATLAS and CMS across various final states. The derived bounds are fundamentally model-dependent, relying on the assumed VLQ-to-SM decay branching ratios, which are fixed based on the VLQ representation. 

Our results reveal exclusion limits that vary across VLQ representations. For singlet configurations, pair production excludes $T$ and $B$ quarks with masses up to approximately 1.48--1.49 TeV. In single production, the mixing parameter $\kappa$, which governs VLQ interactions with SM particles, is constrained to values below 0.26 for $m_T \sim 1.5$ TeV, gradually relaxing to 0.42 near $m_T = 2$ TeV. For doublet representations, pair production extends the exclusion bounds slightly higher, up to 1.49--1.52 TeV, while single production offers additional sensitivity, with $\kappa$ constrained between 0.2 and 0.7, depending on the mass range.

Exotic VLQs such as $X$ and $Y$ exhibit distinctive exclusion patterns. Pair production excludes masses up to 1.46 TeV and 1.7 TeV for $X$ and $Y$, respectively. Single production further constrains $\kappa$, excluding values above 0.16 for $X$ quarks in the mass range $m_X \sim [1.46, 1.6]$ TeV, tightening to $\kappa \leq 0.2$ for $m_X \sim [1.6, 1.8]$ TeV. For $Y$ quarks, single production excludes $\kappa \geq 0.26$ for $m_Y \sim [1.7, 1.8]$ TeV, particularly in the $BY$ doublet configuration, where decays into final states such as $bbbq'$ provide strong sensitivity.

Altogether, these findings underscore the complementarity between QCD-driven pair production and EW-mediated single production in refining exclusion limits. For instance, in the $(T, B)$ doublet configuration, pair production processes $pp \to TT$ and $pp \to BB$ contribute to final states involving leptons, $b$-jets, and MET, while single production enhances sensitivity at higher masses. Similarly, in the $(X, T, B)$ triplet, processes like $pp \to XX$ and $pp \to BB$ produce overlapping intermediate states, strengthening the exclusion bounds through enhanced cross-sections.

Our analysis reflects the evolving experimental constraints on VLQs and their implications for theoretical models. As the LHC dataset grows, the exclusion bounds presented here will continue to improve, paving the way for potential discoveries. To support future studies, the results obtained in this work will soon be accessible through the release of \texttt{VLQBounds}, providing an efficient tool to evaluate VLQ scenarios in light of the latest experimental constraints.

\section{Acknowledgments}
{R.B. is supported in part by the PIFI Grant No. 110200EZ52.}
SM is supported in part through the NExT Institute and the STFC Consolidated Grant ST/L000296/1. QSY is supported by the
Natural Science Foundation of China under the Grants No. 11875260 and No. 12275143.

\appendix

\section{Lagrangian}
\label{sec:a}

The interactions of SM quarks ($t, b$) with VLQs modify the SM Lagrangian as follows:
\begin{eqnarray}
\mathcal{L}_W & = & -\frac{g}{\sqrt 2} \bar t \gm \left( V_{tb}^L P_L + V_{tb}^R P_R \right) b \Wm^+ +\text{H.c.} \,, \notag \\ 
\mathcal{L}_Z & = & -\frac{g}{2 c_W} \bar t \gm \left( X_{tt}^L P_L + X_{tt}^R P_R - 2 Q_t s_W^2 \right) t \Zm \notag \\
& & -\frac{g}{2 c_W} \bar b \gm \left( -X_{bb}^L P_L - X_{bb}^R P_R - 2 Q_b s_W^2 \right) b \Zm \,, \notag \\
\mathcal{L}_H & = & -\frac{g m_t}{2M_W} Y_{tt} \bar t t H -\frac{g m_b}{2M_W} Y_{bb} \bar b b H \,,
\label{ec:ll}
\end{eqnarray}
where $P_{L,R}$ are the projection operators, $Q_t$ and $Q_b$ are the electric charges of $t$ and $b$ quarks, and $s_W$ and $c_W$ denote the sine and cosine of the weak mixing angle, respectively. The charged current couplings $V_{tb}^{L,R}$, the neutral couplings $X_{tt}^{L,R}$, $X_{bb}^{L,R}$, and the Yukawa couplings $Y_{tt}$, $Y_{bb}$ are listed in Tables~\ref{tab:llW}, \ref{tab:llZ}, and \ref{tab:llH}, respectively.

\begin{table}[htb]
	\begin{center}
		\begin{tabular}{c|cc}
			& $V_{tb}^L$ & $V_{tb}^R$ 
			\\ \hline
			\ts & $\clx$ & 0
			\\
			\bs & $\clx$ & 0
			\\
			\xt & $\clx$ & 0
			\\
			\tb & $\clu \cld + \slu \sld $ & $\sru \srd $
			\\
			\by & $\clx$ & 0
			\\
			\xtb & $\clu \cld + \sqt \slu \sld$ & $\sqt \sru \srd$
			\\
			\tby & $\clu \cld + \sqt \slu \sld$ & $\sqt \sru \srd$
		\end{tabular}
		\caption{Light-light couplings to the $W$ boson.}
		\label{tab:llW}
	\end{center}
\end{table}

\begin{table}[htb]
	\begin{center}
		\begin{tabular}{c|cccc}
			& $X_{tt}^L$ & $X_{tt}^R$ & $X_{bb}^L$ & $X_{bb}^R$ 
			\\ \hline
			\ts & $\clx^2$ & 0 & 1 & 0
			\\
			\bs & 1 & 0 & $\clx^2$ & 0
			\\
			\xt & $\clx^2-\slx^2$ & $-\srx^2$ & 1 & 0
			\\
			\tb & 1 & $(\sru)^2$ & 1 & $(\srd)^2$
			\\
			\by & 1 & 0 & $\clx^2-\slx^2$ & $-\slx^2$
			\\
			\xtb & $(\clu)^2$ & 0 & $1+(\sld)^2$ & $2(\srd)^2$
			\\
			\tby & $1+(\slu)^2$ & $2(\sru)^2$ & $(\cld)^2$ & 0
		\end{tabular}
		\caption{Light-light couplings to the $Z$ boson.}
		\label{tab:llZ}
	\end{center}
\end{table}

\begin{table}[htb]
	\begin{center}
		\begin{tabular}{c|cc}
			& $Y_{tt}$ & $Y_{bb}$  
			\\ \hline
			\ts & $\clx^2$ & $1$
			\\
			\bs & $1$ & $\clx^2$
			\\
			\xt & $\crx^2$ & $1$
			\\
			\tb & $(\cru)^2$ & $(\crd)^2$
			\\
			\by & $1$ & $\crx^2$ 
			\\
			\xtb & $(\clu)^2$ & $(\cld)^2$
			\\
			\tby & $(\clu)^2$ & $(\cld)^2$
		\end{tabular}
		\caption{Light-light couplings to the Higgs boson.}
		\label{tab:llH}
	\end{center}
\end{table}


Interactions of the VLQs $Q$ ($Q = X, T, B, Y$) follow an analogous structure:
\begin{eqnarray}
\mathcal{L}_W & = & -\frac{g}{\sqrt 2} \bar Q \gm \left( V_{QQ'}^L P_L + V_{QQ'}^R P_R \right) Q' \Wm^+ +\text{H.c.} \,, \notag \\ 
\mathcal{L}_Z & = & -\frac{g}{2 c_W} \bar Q \gm \left( \pm X_{QQ}^L P_L \pm X_{QQ}^R P_R - 2 Q_Q s_W^2 \right) Q \Zm \,, \notag \\
\mathcal{L}_H & = & -\frac{g m_Q}{2M_W} Y_{QQ} \bar Q Q H \,,
\label{ec:HH}
\end{eqnarray}
where the charge-dependent sign in the $Z$-boson coupling is $+$ for $X, T$ and $-$ for $B, Y$. The corresponding couplings are detailed in Tables~\ref{tab:hhW}, \ref{tab:hhZ}, and \ref{tab:hhH}.

\begin{table}[htb]
	\begin{center}
		\begin{tabular}{c|cccccc}
			& $V_{XT}^L$ & $V_{XT}^R$ & $V_{TB}^L$ & $V_{TB}^R$ & $V_{BY}^L$ & $V_{BY}^R$
			\\ \hline
			\xt & $\clx$ & $\crx$ & -- & -- & -- & --
			\\
			\tb & -- & -- & $\clu \cld + \slu \sld $ & $\cru \crd$ & -- & --
			\\
			\by & -- & -- & -- & -- & $\clx$ & $\crx$
			\\
			\xtb & $\sqt \clu$ & $\sqt \cru$ & $\slu \sld + \sqt \clu \cld$ & $\sqt \cru \crd$ & -- & --
			\\
			\tby & -- & -- & $\slu \sld + \sqt \clu \cld$ & $\sqt \cru \crd$ & $\sqt \cld$ & $\sqt \crd$
		\end{tabular}
		\caption{Heavy-heavy couplings to the $W$ boson.}
		\label{tab:hhW}
	\end{center}
\end{table}

\begin{table}[htb]
	\begin{center}
		\begin{tabular}{c|cccccccc}
			& $X_{XX}^L$ & $X_{XX}^R$ & $X_{TT}^L$ & $X_{TT}^R$ & $X_{BB}^L$ & $X_{BB}^R$ & $X_{YY}^L$ & $X_{YY}^R$
			\\ \hline
			\ts & -- & -- & $\slx^2$ & 0 & -- & -- & -- & --
			\\
			\bs & -- & -- & -- & -- & $\slx^2$ & 0 & -- & --
			\\
			\xt & 1 & 1 & $\slx^2 - \clx^2$ & $-\crx^2$ & -- & -- & -- & --
			\\
			\tb & -- & -- & 1 & $(\cru)^2$ & 1 & $(\crd)^2$ & -- & --
			\\
			\by & -- & -- & -- & -- & $\slx^2 - \clx^2$ & $-\crx^2$ & 1 & 1
			\\
			\xtb & 2 & 2 & $(\slu)^2$ & 0 & $1+(\cld)^2$ & $2 (\crd)^2$ & -- & --
			\\
			\tby & -- & -- & $1+(\clu)^2$ & $2 (\cru)^2$ & $(\sld)^2$ & 0 & 2 & 2
		\end{tabular}
		\caption{Heavy-heavy couplings to the $Z$ boson.}
		\label{tab:hhZ}
	\end{center}
\end{table}

\begin{table}[htb]
	\begin{center}
		\begin{tabular}{c|cccc}
			& $Y_{XX}$ & $Y_{TT}$ & $Y_{BB}$ & $Y_{YY}$ 
			\\ \hline
			\ts & -- & $\slx^2$ & -- & --
			\\
			\bs & -- & -- & $\slx^2$ & --
			\\
			\xt & 0 & $\srx^2$ & -- & --
			\\
			\tb & -- & $(\sru)^2$ & $(\srd)^2$ & --
			\\
			\by & -- & -- & $\srx^2$ & 0
			\\
			\xtb & 0 & $(\slu)^2$ & $(\sld)^2$ & --
			\\
			\tby & -- & $(\slu)^2$ & $(\sld)^2$ & 0
		\end{tabular}
		\caption{Heavy-heavy couplings to the Higgs boson.}
		\label{tab:hhH}
	\end{center}
\end{table}


The interactions between heavy ($Q$) and light ($q$) quarks are given by:
\begin{eqnarray}
\mathcal{L}_W & = & -\frac{g}{\sqrt 2} \bar Q \gm \left( V_{Qq}^L P_L + V_{Qq}^R P_R \right) q \Wm^+ +\text{H.c.} \notag \\
& & -\frac{g}{\sqrt 2} \bar q \gm \left( V_{qQ}^L P_L + V_{qQq}^R P_R \right) Q \Wm^+ +\text{H.c.} \,, \notag \\ 
\mathcal{L}_Z & = & -\frac{g}{2 c_W} \bar q \gm \left( \pm X_{qQ}^L P_L \pm X_{qQ}^R P_R \right) Q \Zm +\text{H.c.} \,, \notag \\
\mathcal{L}_H & = & -\frac{g m_Q}{2M_W} \bar q \left( Y_{qQ}^L P_L +  Y_{qQ}^R P_R \right) Q H +\text{H.c.} \,,
\end{eqnarray}

\begin{table}[htb]
	\begin{center}
		\begin{tabular}{c|cccc}
			& $V_{Xt}^L$ & $V_{Xt}^R$ & $V_{Tb}^L$ & $V_{Tb}^R$ 
			\\ \hline
			\ts & -- & -- & $\slx $ & 0 
			\\
			\xt & $-\slx $ & $-\srx $ & $\slx $ & 0 
			\\
			\tb & -- & -- & $\slu \cld e^{-i \phi_u} - \clu \sld e^{-i \phi_d}$ & $-\cru \srd e^{-i \phi_d}$
			\\
			\xtb & $-\sqt \slu $ & $-\sqt \sru $ & $(\slu \cld - \sqt \clu \sld) $ & $-\sqt \cru \srd $
			\\
			\tby & -- & -- & $(\slu \cld - \sqt \clu \sld) $ & $-\sqt \cru \srd $
		\end{tabular}
		\caption{Heavy-light couplings to the $W$ boson.}
		\label{tab:WHl}
	\end{center}
\end{table}

\begin{table}[htb]
	\begin{center}
		\begin{tabular}{c|cccc}
			& $V_{tB}^L$ & $V_{tB}^R$ & $V_{bY}^L$ & $V_{bY}^R$
			\\ \hline
			\bs & $\slx $ & 0 & -- & --
			\\
			\tb & $\clu \sld -\slu \cld $ & $-\sru \crd $ & -- & --
			\\
			\by & $\slx $ & 0 & $-\slx $ & $-\srx $
			\\
			\xtb & $(\clu \sld - \sqt \slu \cld) $ & $-\sqt \sru \crd $ & -- & --
			\\
			\tby & $(\clu \sld - \sqt \slu \cld)$ & $-\sqt \sru \crd $ & $-\sqt \sld $ & $-\sqt \srd $
		\end{tabular}
		\caption{Light-heavy couplings to the $W$ boson.}
		\label{tab:WlH}
	\end{center}
\end{table}

\begin{table}[htb]
	\begin{center}
		\begin{tabular}{c|cccc}
			& $X_{tT}^L$ & $X_{tT}^R$ & $X_{bB}^L$ & $X_{bB}^R$ 
			\\ \hline
			\ts & $\slx \clx $ & 0 & -- & --
			\\
			\bs & -- & -- & $\slx \clx $ & 0
			\\
			\xt & $2 \slx \clx $ & $\srx \crx $ & -- & --
			\\
			\tb & 0 & $-\sru \cru $ & 0 & $-\srd \crd $
			\\
			\by & -- & -- & $2\slx \clx $ & $\srx \crx $ 
			\\
			\xtb & $\slu \clu $ & 0 & $-\sld \cld $ & $-2\srd \crd $
			\\
			\tby & $-\slu \clu $ & $-2\sru \cru $ & $\sld \cld $ & 0 
		\end{tabular}
		\caption{Light-heavy couplings to the $Z$ boson.}
		\label{tab:ZlH}
	\end{center}
\end{table}

\clearpage

\begin{table}[htb]
	\begin{center}
		\begin{tabular}{c|cccc}
			& $Y_{tT}^L$ & $Y_{tT}^R$ & $Y_{bB}^L$ & $Y_{bB}^R$  
			\\ \hline
			\ts & $\frac{m_t}{m_T} \slx \clx $ & $\slx \clx $ & -- & --
			\\
			\bs & -- & -- & $\frac{m_b}{m_B} \slx \clx $ & $\slx \clx $
			\\
			\xt & $\srx \crx $ & $\frac{m_t}{m_T} \srx \crx $ & -- & --
			\\
			\tb & $\sru \cru $ & $\frac{m_t}{m_T} \sru \cru $ & $\srd \crd $ & $\frac{m_b}{m_B} \srd \crd $
			\\
			\by & -- & -- & $\srx \crx $ & $\frac{m_b}{m_B} \srx \crx $ 
			\\
			\xtb & $\frac{m_t}{m_T} \slu \clu $  & $\slu \clu $ & $\frac{m_b}{m_B} \sld \cld $ & $\sld \cld $
			\\
			\tby & $\frac{m_t}{m_T} \slu \clu $ & $\slu \clu $ & $\frac{m_b}{m_B} \sld \cld $ & $\sld \cld $
		\end{tabular}
		\caption{Light-heavy couplings to the Higgs boson.}
		\label{tab:HlH}
	\end{center}
\end{table}

\section{Heavy quark decay widths}
\label{sec:b}

The decay widths of $X$ and $Y$ VLQs are calculated as follows:
\begin{align}
\Gamma(X \to W^+ t) & = \frac{g^2}{64 \pi}  \frac{m_X}{M_W^2} \lambda(m_X,m_t,M_W)^{1/2} \left\{
(|V_{Xt}^L|^2+|V_{Xt}^R|^2)  \right. \nonumber \\
& \left. \times \left[ 1+r_W^2-2 r_t^2
-2 r_W^4  + r_t^4 +r_W^2 r_t^2
\right]  -12 r_W^2 r_t \RE V_{Xt}^L V_{Xt}^{R*} \right\} \,, \notag \\
\Gamma(Y \to W^- b) & = \frac{g^2}{64 \pi}  \frac{m_T}{M_W^2} \lambda(m_Y,m_b,M_W)^{1/2} \left\{
(|V_{bY}^L|^2+|V_{bY}^R|^2)  \right. \nonumber \\
& \left. \times \left[ 1+r_W^2-2 r_b^2
-2 r_W^4  + r_b^4 +r_W^2 r_b^2
\right]  -12 r_W^2 r_b \RE V_{bY}^L V_{bY}^{R*} \right\}
\,.
\label{ec:GammaXY}
\end{align}
where $r_x \equiv m_x / m_Q$, where $x=t,b,W,Z,H$ and $Q$ is the heavy quark and
\begin{equation}
\lambda(x,y,z) \equiv (x^4 + y^4 + z^4 - 2 x^2 y^2 
- 2 x^2 z^2 - 2 y^2 z^2) \,.
\end{equation}%
The charged current mixings $V$ are given in Tabs.~\ref{tab:WlH} and~\ref{tab:WHl} of Appendix~\ref{sec:a}.
The partial widths for $T$ decays, including all possible mixing terms, are
\begin{align}
\Gamma(T \to W^+ b) & = \frac{g^2}{64 \pi}  \frac{m_T}{M_W^2} \lambda(m_T,m_b,M_W)^{1/2} \left\{
(|V_{Tb}^L|^2+|V_{Tb}^R|^2)  \right. \nonumber \\
& \left. \times \left[ 1+r_W^2-2 r_b^2 -2 r_W^4  + r_b^4 +r_W^2 r_b^2
\right]  -12 r_W^2 r_b \RE V_{Tb}^L V_{Tb}^{R*} \right\}
\,, \nonumber \\
\Gamma(T \to Z t) & = \frac{g}{128 \pi c_W^2}  \frac{m_T}{M_Z^2} \lambda(m_T,m_t,M_Z)^{1/2}
\left\{ (|X_{tT}^L|^2 + |X_{tT}^R|^2) \right. \nonumber \\
& \left. \times  \left[ 1 + r_Z^2 - 2  r_t^2 - 2  r_Z^4  + r_t^4
+ r_Z^2 r_t^2 \right]  -12 r_Z^2 r_t \RE X_{tT}^L X_{tT}^{R*}  \right\} \,, \nonumber \\
\Gamma(T \to H t) & = \frac{g^2}{128 \pi}
\frac{m_T}{M_W^2} \lambda(m_T,m_t,M_H)^{1/2} |Y_{tT}|^2 \left[ 1 + 6 r_t^2 - r_H^2 
+ r_t^4 - r_t^2 r_H^2 \right] \,,
\label{ec:GammaT}
\end{align}
For the $B$ quark, the expressions for decay widths are fully analogous,
\begin{align}
\Gamma(B \to W^- t) & = \frac{g^2}{64 \pi}  \frac{m_B}{M_W^2} \lambda(m_B,m_t,M_W)^{1/2} \left\{
(|V_{tB}^L|^2+|V_{tB}^R|^2)  \right. \nonumber \\
& \left. \times \left[ 1+r_W^2-2 r_t^2  -2 r_W^4  + r_t^4 +r_W^2 r_t^2
\right]  -12 r_W^2 r_t \RE V_{tB}^L V_{tB}^{R*} \right\}
\,, \nonumber \\
\Gamma(B \to Z b) & = \frac{g}{128 \pi c_W^2}  \frac{m_B}{M_Z^2} \lambda(m_B,m_b,M_Z)^{1/2}
\left\{ (|X_{bB}^L|^2 + |X_{bB}^R|^2) \right. \nonumber \\
& \left. \times  \left[ 1 + r_Z^2 - 2  r_b^2 - 2  r_Z^4  + r_b^4
+ r_Z^2 r_b^2 \right]  -12 r_Z^2 r_b \RE X_{bB}^L X_{bB}^{R*}  \right\} \,, \nonumber \\
\Gamma(B \to H b) & = \frac{g^2}{128 \pi}
\frac{m_B}{M_W^2} \lambda(m_B,m_b,M_H)^{1/2} |Y_{bB}|^2 \left[ 1 + 6 r_b^2 - r_H^2 
+ r_b^4 - r_b^2 r_H^2 \right] \,.
\label{ec:GammaB}
\end{align}
The neutral current mixing parameters, $X$, are provided in Table~\ref{tab:ZlH} of Appendix~\ref{sec:a}. For partial decay widths to Higgs final states, $Y_{qQ}$ represents the dominant light-heavy Yukawa coupling listed in Table~\ref{tab:HlH}, specifically $Y_{qQ}^R$ for singlets and triplets, and $Y_{qQ}^L$ for doublets.

\section{Relationship between $\kappa$ and Mixing Angles}\label{apB}

In experimental searches, limits on VLQs are commonly expressed using different parameter notations. This work adopts the framework established in~\cite{Aguilar-Saavedra:2013qpa}, widely used in theoretical studies. However, certain LHC analyses, notably those by CMS, employ the simplified parameterization presented in~\cite{Buchkremer:2013bha} or~\cite{Fuks:2016ftf}. This approach introduces the parameter $\kappa$ to represent the coupling strength of VLQs to third-generation Standard Model quarks. To ensure consistency across representations when interpreting experimental results, Tab.~\ref{tab:kappa} provides the relationship between $\kappa$ and the mixing angles relevant to each VLQ configuration.

\begin{table}[h!]
	\centering
	\renewcommand{\arraystretch}{1.2}
	\begin{tabular}{c|c|c}
\toprule
		\textbf{VLQ} & \textbf{Representation} & \textbf{$\kappa$ Expression in terms of mixing angles} \\
\toprule
		\multirow{3}{*}{$T$ } & Singlet & $ \sqrt{2} \, s_L$ \\
		& Doublet ($X, T$) & $ s_R \, c_R$ \\
		& Doublet ($T, B$) & $ s^u_R \, c^u_R$ \\
		\hline
		\multirow{3}{*}{$B$ } & Singlet & $\sqrt{2} \, s_L$ \\
		& Doublet ($B, Y$) & $ s_R \, c_R$ \\
		& Doublet ($T, B$) & $ s^d_R \, c^d_R$ \\
		\hline
		$X$  & Doublet ($X, T$) & $ s_R$ \\
					\hline
		$Y$  & Doublet ($B, Y$) & $ s_R$ \\
		\hline
		$Y$ & Triplet ($T, B, Y$) & $ \sqrt{2} \, s^d_L$ \\
	
\bottomrule
	\end{tabular}
	\caption{Translation of $\kappa$ in terms of mixing angles for various VLQ representations.}
	\label{tab:kappa}
\end{table}

\newpage
\bibliography{main.bib} 
\end{document}